\documentclass[a4paper,useAMS,fleqn,usenatbib]{mnras}

\usepackage[T1]{fontenc}
\usepackage{ae,aecompl}

\usepackage{footnote,graphicx,natbib,color,multirow,amsmath,url}

\definecolor{titlecol}{rgb}{0,0,1}
\definecolor{titlecol2}{rgb}{0,0.4,0}
\definecolor{titlecol3}{rgb}{0,0.5,0.8}

\def\changed    {}

\def\oiii		{$\mathrm{\left[ O \textsc{iii}\right] }$}

\def\nii		{$\mathrm{\left[ N \textsc{ii}\right] }$}

\def\feii		{$\mathrm{Fe~\textsc{ii} }$}

\def\gandalf     {{\tt GANDALF}}
\def\emcee      {{\tt emcee}}
\def\diskdom    {\textsc{diskdom}}

\def\qsocontrol    {\textsc{qsocontrol}}

\def\msun	       {$\rm{M}_{\odot}$}
\def\mmsun	{\rm{M}_{\odot}}

\def\ha               {H$\alpha$}

\def\mbh           {$M_{\rm BH}$}
\def\mmbh        {M_{\rm BH}}
\def\lbol             {$L_{\rm bol}$}
\def\mlbol          {L_{\rm bol}}
\def\ledd            {$L_{\rm Edd}$}
\def\mledd         {L_{\rm Edd}}
\def\eddrat        {$\lambda_{\rm Edd}$}
\def\meddrat     {\lambda_{\rm Edd}}
\def\mdot          {$\dot m$}
\def\mmdot       {\dot m}

\def\lesssim{\mathrel{\hbox{\rlap{\hbox{\lower3pt\hbox{$\sim$}}}\hbox{\raise2pt\hbox{$<$}}}}}
\def\gtrsim{\mathrel{\hbox{\rlap{\hbox{\lower3pt\hbox{$\sim$}}}\hbox{\raise2pt\hbox{$>$}}}}}

\begin{document}


\title[SMBHs co-evolve with their host disk galaxies]{Supermassive black holes in disk-dominated galaxies outgrow their bulges and co-evolve with their host galaxies}

\author[Simmons, Smethurst \& Lintott]{
B. D. Simmons$^{1, 2, 3}$\thanks{Einstein Fellow}\thanks{E-mail: bdsimmons@ucsd.edu},
R. J. Smethurst$^{2, 4}$
and C. Lintott$^{2}$
\\
$^{1}$Center for Astrophysics and Space Sciences (CASS), Department of Physics, University of California, San Diego, CA 92093, USA\\
$^{2}$Oxford Astrophysics, Denys Wilkinson Building, Keble Road, Oxford OX1 3RH, UK\\
$^{3}$Balliol College, Oxford OX1 3BJ, UK\\
$^{4}$School of Physics \& Astronomy, The University of Nottingham, University Park, Nottingham, NG7 2RD, UK
}

\date{Accepted 2017 May 26. Received 2017 May 20; in original form 2017 March 14}

\pubyear{2017}


\label{firstpage}
\pagerange{\pageref{firstpage}--\pageref{lastpage}}
\maketitle

%
%
%
%
%
%

\begin{abstract}

  The deep connection between galaxies and their supermassive black holes is central to modern astrophysics and cosmology. The observed correlation between galaxy and black hole mass is usually attributed to the contribution of major mergers to both. We make use of a sample of galaxies whose disk-dominated morphologies indicate a major-merger-free history and show that such systems are capable of growing supermassive black holes at rates similar to quasars. 
Comparing black hole masses to conservative upper limits on bulge masses, we show that the black holes in the sample
are typically larger than expected if processes creating bulges are also the primary driver of black hole growth.
The same relation between black hole and total stellar mass of the galaxy is found for the merger-free sample as for a sample which has experienced substantial mergers, indicating that major mergers do not play a significant role in controlling the coevolution of galaxies and black holes. We suggest that more fundamental processes which contribute to galaxy assembly are also responsible for black hole growth.

\end{abstract}
  
\begin{keywords}
  galaxies: general 
  --- 
  galaxies: evolution
  --- 
  galaxies: active 
  --- 
  galaxies: quasars: supermassive black holes
  --- 
  galaxies: bulges
  --- 
  galaxies: spiral
\end{keywords}

%
%
\section{Introduction}\label{sec:intro}
%
%

Galaxy mergers are a key means of transferring stars from rotation-supported disks to dispersion-supported bulges and ellipticals \citep{toomre77,walker96,hopkins11c,welker15}. While gas-rich galaxies may re-form a disk following a major merger event \citep[for theoretical work see][and for observations of early-types containing disks see \citealt{emsellem11}]{hopkins09c}, the bulge persists as an artifact of the past merger. The minimum galaxy mass ratio required for a merger to create a bulge from a disk is often cited as $1:10$, although recent theoretical studies have introduced some uncertainty into this number \citep{hopkins09c,brook12,kannan15}
and recent work suggests bulges may grow further via means other than mergers \citep{bell17}.
 Nevertheless, galaxies which are clearly disk-dominated have likely had an extremely calm baryon accretion history, evolving in the absence of major mergers since $z \sim 2$ \citep{martig12}.

Across a wide span of redshift and galaxy masses, the strong correlations observed between supermassive black hole mass and galaxy properties (such as velocity dispersion, \citealt{magorrian98,ferrarese00,kormendy01,mcconnell13}; bulge stellar mass, \citealt{marconi03, haringrix04}; and total stellar mass, \citealt{cisternas11,marleau13}) suggest that galaxies co-evolve with their central supermassive black holes \citep[SMBHs; for a review see][]{kormendy13}. These correlations have often been interpreted as an outcome of galaxy-galaxy mergers that grow both bulges and central black holes \citep[e.g.,][]{sanders88,croton06,hopkins06b,peng07}. 

\clearpage

Indeed, at least some populations of luminous quasars are preferentially hosted in ongoing mergers \citep[][although see, e.g., \citealt{dunlop03,mechtley16}]{urrutia08,glikman15,trakhtenbrot16b}, with detailed follow-up of individual objects showing evidence of AGN triggering due to a major merger \citep[e.g.,][]{bessiere14} and population studies indicating that AGN activity peaks in the post-merger phase \citep{ellison13}. Those SMBHs which have grown to the highest masses are generally found in early-type galaxies \citep[e.g.,][]{mcconnell13} with dynamical and morphological configurations indicating a history dominated by major mergers.

If major mergers drive the observed black hole-galaxy correlations, we should expect that the properties of black holes will correlate only with galaxy properties tied to merger histories (i.e., dispersion-supported bulges), and not with properties that grow in the absence of mergers. Some past studies support this hypothesis \citep{kormendy11a,kormendy11b}. However, moderate-luminosity AGN, which represent more typical rates of black hole growth \citep{hasinger05}, are hosted in galaxy populations with high disk-dominated fractions \citep{simmons11,simmons12b,schawinski11a,schawinski12,kocevski12} and that show no increase in merger incidence compared to control samples \citep{grogin05,gabor09,kocevski12}. Within these samples, black holes correlate with total galaxy stellar mass \citep{cisternas11}, i.e., including the mass of the disk.

Merger-free black hole-galaxy evolution has been proposed previously \citep[e.g.,][]{greene10b,jiang11b,cisternas11,simmons13}, and there is some evidence that merger-free processes such as violent disk instabilities (e.g., clump instabilities, \citealt{schawinski11b,bournaud12}; and galactic bars, \citealt{knapen00,hao09,oh12,galloway15}) or purely calm, ``secular'' processes \citep{kormendy04} may be able to grow SMBHs to typical masses in the absence of significant mergers \citep{simmons13}. Owing to the relative rarity of merger-free galaxies hosting growing black holes, however, previously studied samples of bulgeless AGN host galaxies have either been too small to statistically constrain black hole-galaxy relations in the merger-free regime, use selection techniques that preclude sampling the full parameter space of galaxy and black hole masses \citep[e.g.][]{jiang11a,jiang11b,mathur12}, or use obscured AGN, for which the only widely available black hole mass estimation techniques are highly uncertain compared to broad-line estimation methods \citep{marleau13}.

In this work we present a sample of 101 luminous, unobscured AGN hosted in disk-dominated galaxies with stellar masses up to $\sim 2 \times 10^{11} \mmsun $. Using broad \ha\ line emission to estimate black hole masses, we compare black hole masses with total stellar masses and conservative upper limits on bulge stellar masses. We additionally compare derived relations between these quantities with the relations previously observed for bulge-dominated and elliptical galaxies. Similarities \emph{or} differences between black hole-galaxy correlations in bulge-dominated galaxies (with major-merger driven evolutionary histories) and disk-dominated galaxies (with histories dominated by merger-free processes) have important implications for understanding what baryonic and dynamical processes, if any, drive the observed correlations between galaxies and black holes.

In Section \ref{sec:data} we describe the observational data and the selection of disk-dominated galaxies hosting luminous AGN. In Section \ref{sec:masses} we describe the calculation of black hole masses, galaxy stellar masses, and bulge masses. We discuss luminosities and growth rates of AGN in disk-dominated galaxies in Section \ref{sec:lum}, and present galaxy-black hole relations for the sample in Section \ref{sec:bhmassrelations}. Throughout
this paper all cross-matched catalogues use the nearest positional match within $3^{\prime\prime}$, we use the AB magnitude system, and where necessary we adopt a cosmology consistent with $\Lambda$CDM, with $H_{\rm 0}=70~{\rm
km~s^{-1}}$Mpc$^{\rm -1}$, $\Omega_{\rm m}=0.3$ and $\Omega_{\rm \Lambda}=0.7$ \citep{bennett13}.

%
%
\section{Observational Data}\label{sec:data}
%
%

The goal of this study is to investigate black hole growth in galaxies whose growth histories have been dominated by relatively calm processes. We therefore require a sample of growing black holes hosted in disk-dominated galaxies. Optimally, the AGN should have broad emission lines to facilitate measurement of black hole masses via well-established relations between line flux and width and black hole masses. Below we describe the method and data sets used to select a sample of unobscured AGN hosted in disk-dominated galaxies.

\subsection{AGN Selection}

Examining the black hole-galaxy relation requires selection of a sample of unobscured AGN with broad emission lines, so that black hole masses may be estimated from well-established correlations between emission line properties and black hole masses \citep[e.g.,][]{gh06,gh07a,xiao11}. 

Unobscured AGN have characteristic colours in multi-wavelength imaging, particularly in X-ray, optical and infrared bands \citep[e.g.,][]{richards02,bauer04,dstern05}. Given the existence of all-sky surveys at many of the wavelengths relevant to the selection of unobscured AGN, it is now possible to construct samples of sources identified as unobscured AGN with high likelihood.

We select an initial sample of AGN using the W2R sample of \citet{edelson12}, comprised of 4,316 sources identified using multi-wavelength data from the \emph{Wide-field Infrared Survey Explorer} \citep[\emph{WISE};][]{wright10}, Two Micron All-Sky Survey \citep[2MASS;][]{skrutskie06}, and \emph{ROSAT} all-sky survey \citep[RASS;][]{voges99}. This is a photometric, all-sky selection, which combines infrared colours with X-ray information to select unobscured AGN at $\gtrsim 95$ per cent confidence \citep{edelson12}.

\subsection{Selecting disk-dominated AGN host galaxies}\label{sec:select}

\begin{figure*}
\includegraphics[height=220mm]{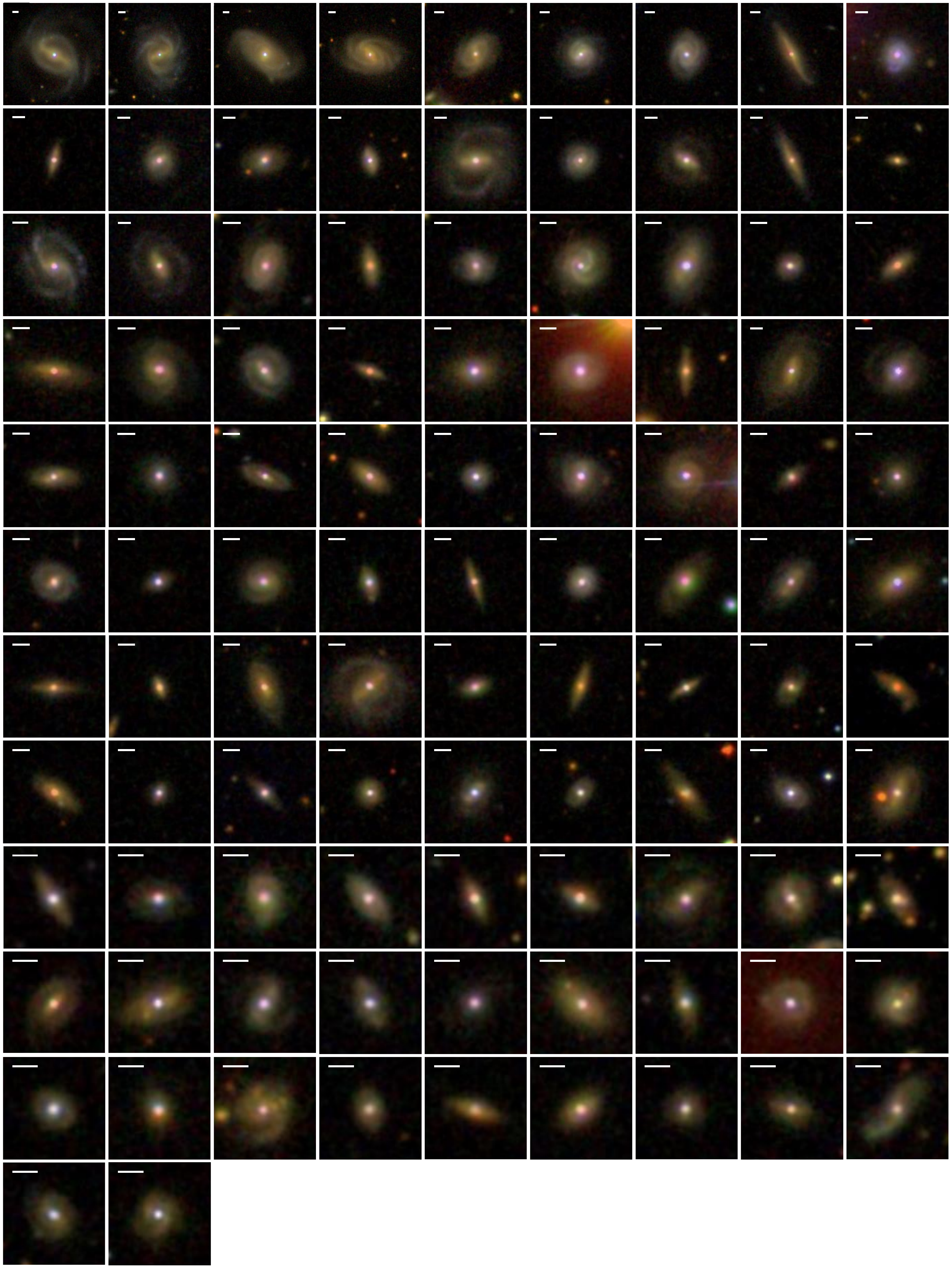}
\caption{Postage stamp SDSS $gri$ images of the sample of 101 disk-dominated galaxies with unobscured AGN, described in Sections \ref{sec:select} and \ref{sec:spectra}. Images are sorted by redshift (reading from top left to bottom right). Scale bars on each panel show $5^{\prime \prime}$. These 101 sources comprise the \diskdom\ sample.
}
\label{fig:exampleimages}
\end{figure*}

Following the AGN selection described above, we further sub-select galaxies imaged by the Sloan Digital Sky Survey. There are 1,844 W2R sources with positional matches having reported coordinates within $3^{\prime \prime}$ of a source in SDSS \citep{york00} Data Release 8 \citep{aihara11}, a fraction consistent with the fractional area of the SDSS versus an all-sky catalog. 76 per cent of this sub-sample has measured redshifts, with a peak redshift distribution of $z \approx 0.12$.  90 percent of sources with redshifts have $z < 0.6$; the distribution has a long tail to $z_{\rm max} = 2.35$.

Using the SDSS colour images, a morphological selection was then performed by a single expert classifier (BDS), who identified 137 systems where disk features such as spiral arms or bars were visible but where no obvious bulge was seen. Edge-on disks without obvious bulges were also included. Given the presence of the AGN, this selection essentially places a limit on the possible bulge mass given the PSF of the central point source (see limits on bulge mass in Section \ref{sec:galmass}).

\subsection{Spectra}\label{sec:spectra}

Of the disk-dominated AGN host galaxies covered by SDSS imaging, 96 have spectra from SDSS Data Release 9 \citep{ahn12}, which includes spectra within the SDSS DR7 \citep{abazajian09}, where spectroscopic coverage is highly complete for galaxies at these magnitudes. The remaining galaxies were not observed due to the low and non-uniform spectral completeness of galaxies in SDSS-III. 
We obtained spectra of the \ha\ emission region for 5 additional sources using the Intermediate Dispersion Spectrograph on the Isaac Newton Telescope (INT) from 21st-23rd May 2014. These spectra were reduced using the standard reduction pipeline. All 101 spectra show broadened \ha\ line emission indicating unobscured AGN. 

\begin{figure*}
\includegraphics[width=\textwidth]{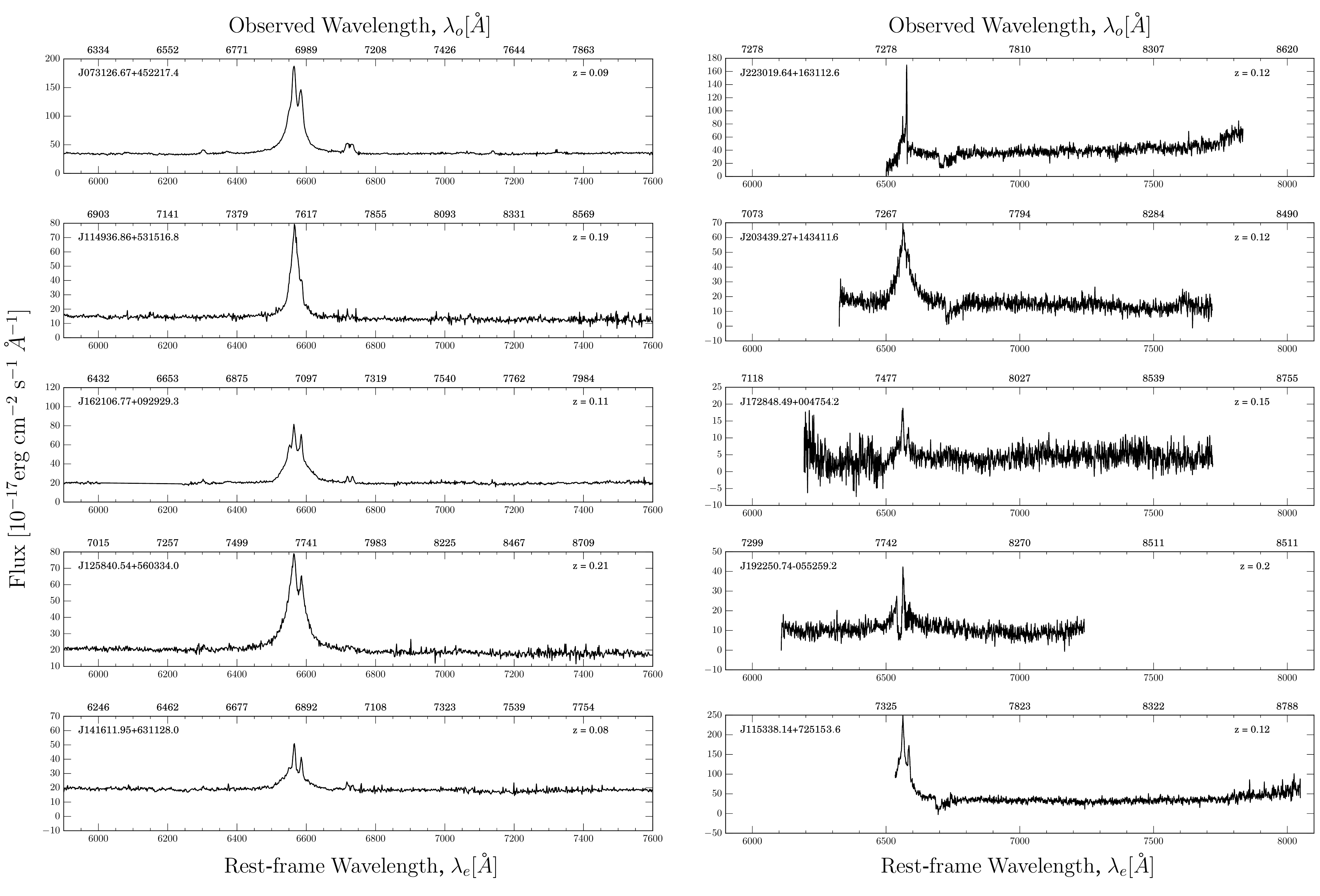}
\caption{
Example spectra from the \diskdom\ sample. The left column shows 5 randomly-selected SDSS spectra (of 96 total) from the sample, zoomed to show the \ha\ region of the spectrum. The right column shows the INT spectra for the 5 galaxies observed using the Intermediate Dispersion Spectrograph. Each panel shows the same rest-frame wavelength range (bottom axis of each panel); observed wavelengths are shown on the top axis of each panel, with redshifts in the top right of each panel. All spectra show broadened \ha\ emission, confirming that the multi-wavelength AGN selection employed here efficiently selects unobscured AGN.
}
\label{fig:examplespectra}
\end{figure*}

\begin{figure}
\includegraphics[width=0.5\textwidth]{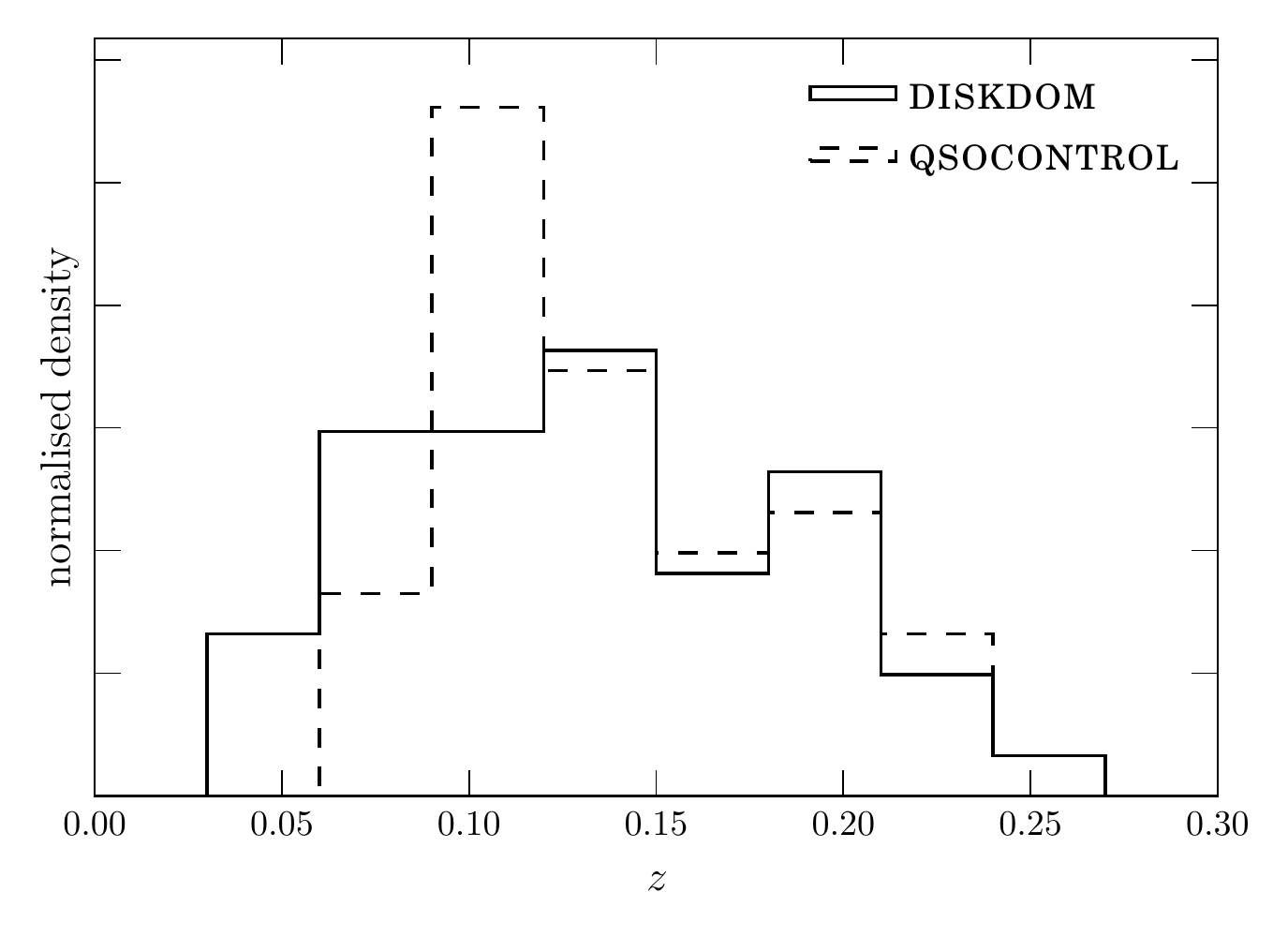}
\caption{Redshift distribution for all 101 sources (solid) in the \diskdom\ sample (Section \ref{sec:spectra}). The dashed line shows the redshift distribution of the \qsocontrol\ sample (Section \ref{sec:sdssqso}). Despite the slight differences in distributions at the low-redshift end, a K-S test cannot reject the null hypothesis that the distributions are drawn from the same parent sample ($p = 0.1$).
}
\label{fig:redshifts}
\end{figure}

Figure \ref{fig:exampleimages} shows SDSS cutouts of these galaxies; Figure \ref{fig:examplespectra} shows the INT spectra along with 5 example SDSS spectra selected at random from the full sample. The mean redshift of the sample is $\left< z \right> = 0.132$, with the highest-redshift source having $z = 0.244$. Figure \ref{fig:redshifts} shows the redshift distribution of the sample. These 101 disk-dominated galaxies hosting luminous, broad-line AGN comprise the primary sample in this work, which is hereafter referred to as the \diskdom\ sample.

\subsection{A comparison sample of SDSS quasars}\label{sec:sdssqso}

We select a comparison sample of unobscured AGN observed at the same epochs as the \diskdom\ sample in order to place the AGN properties of the \diskdom\ sample into context. Specifically, we select AGN from the full catalog of 105,783 SDSS quasars with black hole mass measurements from \citet{shen11}. This catalog includes 38 of the AGN in the \diskdom\ sample; we discuss these further in section in Section \ref{sec:bhmass} below. After removing these 38 sources from the full Shen et al. sample, we select a redshift-matched subsample of 101 quasars. We refer to this sample throughout the text as the \qsocontrol\ sample, and show their redshift distribution for comparison to the \diskdom\ sample in Figure \ref{fig:redshifts}. 

There are no Shen et al. quasars with $z < 0.0645$ that are not already in the \diskdom\ sample. As the lowest-redshift source in the \diskdom\ sample has $z = 0.031$, the \qsocontrol\ sample has a slightly different redshift distribution at $z < 0.1$. Nevertheless, the full distributions are similar: a Kolmogorov-Smirnov test indicates the null hypothesis cannot be ruled out ($p = 0.1$). We have additionally verified that the statistically significant differences in sample distributions discussed in Section \ref{sec:lum} persist even when only considering sub-samples of the \diskdom\ and \qsocontrol\ samples that exclude the low-redshift end of both distributions.

We note that the host galaxies of the \qsocontrol\ sample are in general not disk-dominated: all 54 galaxies from the \qsocontrol\ sample that also have morphological classifications from Galaxy Zoo 1 \citep{lintott08,lintott11} have $p_{CS} < 0.5$, where $p_{CS}$ is the percentage of votes for a combined spiral or edge-on classification (corrected for the effects of loss of resolution due to redshift effects). That is, for every galaxy with a Galaxy Zoo morphological classification in the \qsocontrol\ sample, fewer than half of classifiers indicated the galaxy had a disk. The mean combined spiral vote percentage for the sample is $\langle p_{CS}\rangle = 0.14$. In contrast, the mean combined spiral vote percentage for the 83 galaxies in the \diskdom\ sample with Galaxy Zoo 1 classifications is $\langle p_{CS}\rangle = 0.54$. We reject the null hypothesis that the $p_{CS}$ distributions are drawn from the same parent distribution (K-S $p = 3 \times 10^{-14}$). The AGN in the \qsocontrol\ sample are at similar redshifts to the \diskdom\ sample, but their host galaxies have very different morphologies.

%
%
\section{Stellar and Black Hole Masses}\label{sec:masses}
%
%

In order to investigate how galaxy and black hole growth may correlate in the absence of major mergers, we require accurate estimates of black hole masses and galaxy stellar masses. Briefly, we estimate black hole masses from broad optical emission line width and luminosity; we compute stellar masses using host galaxy colours and luminosities. We also derive upper limits to bulge stellar masses. Below we describe each of these methods in further detail.

\subsection{Galaxy and bulge stellar masses}\label{sec:galmass}

We calculate stellar masses using the well-studied relation between stellar mass, absolute galaxy $r$-band magnitude, and $u-r$ galaxy colour \citep[corrected for galactic extinction;][]{schlegel98}\footnote{\url{https://github.com/rjsmethurst/ebvpy}}, following the method of \citet{baldry06}. We remove the AGN contribution to the luminosity and colour of each galaxy by subtracting the flux in the SDSS {\tt psfMag} from the flux in {\tt modelMag}. {\tt psfMag} is the best estimate of unresolved emission, while {\tt modelMag} is the optimal quantity for computing aperture-matched source colours\footnote{\url{https://www.sdss3.org/dr10/algorithms/magnitudes.php}}. 

The full sample of total stellar masses spans the range from $8 \times 10^9 \mmsun < M_* < 2 \times 10^{11} \mmsun $, with a mean galaxy stellar mass of $4 \times 10^{10} \mmsun $. Each individual mass has an uncertainty of  $0.3$ dex from the scatter in the colour-luminosity relation.

The nuclear emission, as estimated via comparison of {\tt psfMag} to {\tt modelMag}, is generally between 20 to 200 per cent of the galaxy-only emission. The presence of the luminous AGN thus severely compromises the estimates of the bulge-to-total ratio in the host galaxy provided by, e.g., the {\tt fracDeV} parameter reported in the SDSS catalogs. 80 per cent of the galaxies in this sample have ${\tt fracDeV}=1$ in the $r$-band, a measurement that usually represents a pure \citet{devaucouleurs} bulge. However, all galaxies in the sample show clear visual signatures of dominant disks (Figure \ref{fig:exampleimages}). None of the photometric parameters measured by the SDSS pipeline allow for the dual presence of an AGN and a host galaxy; without such considerations the unresolved AGN light is likely to be attributed to the compact bulge component in a bulge-disk model fit \citep{simmons08,kim08,koss11}.

AGN-host decomposition based on 2-dimensional image fitting \citep[e.g.,][]{simard98,peng02,haussler07,peng10} is more reliable \citep[e.g.][]{mclure99,urry00,mclure01,sanchez04,pierce07,gabor09,simmons11,simmons12b,simmons13,koss11,schawinski12}, although even in high-resolution \emph{Hubble Space Telescope (HST)} images \citep{simmons08,kim08} or SDSS imaging at $z \gtrsim 0.06$ \citep{koss11,simmons13} the recovered bulge-to-total ratio can be highly uncertain, particularly for disk-dominated galaxies with a small bulge or pseudo-bulge component. In the majority of sources in this sample, the nuclear emission is too bright and the image resolution too low for a reliable bulge-to-disk decomposition.

Nevertheless, constraining the bulge-black hole mass relation in these galaxies is possible even with upper limits to the bulge contributions to the host galaxies. We can obtain such limits using existing structural parameters from large-scale studies performing bulge-disk decompositions of SDSS galaxies. While such studies do not account for the presence of an AGN, their tendency to overestimate the bulge-to-total ratio as a result means that bulge masses derived from these quantities may be taken to be conservative upper limits.

We use the bulge-disk decompositions of \citet{simard11}, who fit multiple models to 1.12 million galaxies in the SDSS DR7 Legacy area and determined best-fit models and structural parameters for each. We take the $r$-band bulge-to-total ratio of the best-fit model as an upper limit to the true bulge-to-total ratio of these AGN host galaxies. To convert limits on bulge luminosities to limits on bulge masses, we assume the mass-to-light ratio of the bulge is equal to the mass-to-light ratio of the disk. This is a reasonable assumption in disk-dominated galaxies, where many of the ``bulge'' components are likely to be rotationally-supported pseudo-bulges \citep{kormendy04} with stellar populations similar to that of the disk \citep[e.g.,][]{graham01a,simmons13}.

The upper bulge-to-total limits of the 90 galaxies in the \diskdom\ sample which were included in the \citeauthor{simard11} study lie between $0.13 \leq \left({\rm B/Tot}\right)_{\rm max} \leq 1.0$, with a mean value of 0.5. Applying these bulge-to-total limits to the stellar masses computed using the colour-luminosity relation \citep{bell01,baldry06} results in bulge stellar mass upper limits of $3 \times 10^9~\mmsun < {\rm M_{bulge}} < 7 \times 10^{10}~\mmsun $. With AGN-host fitting, we additionally constrained 2 of 5 INT sources to have bulge-to-total ratios of $0.3 \pm 0.2$ and $0.47 \pm 0.2$. For the 9 galaxies in the \diskdom\ sample without bulge mass constraints from bulge-disk decompositions, we assume an upper bulge fraction limit of 1.0, and note that excluding these galaxies does not substantially alter our results. We compare these mass limits with black hole masses for the sample in Section \ref{sec:bhmassrelations}.

\subsection{Black hole mass estimates}\label{sec:bhmass}

The selection of unobscured AGN facilitates accurate estimates of black hole masses based on single-epoch spectra with broadened lines due to emission originating from within the black hole sphere of influence. We use the established relation between black hole mass and the FWHM and luminosity in the broad \ha\ line. Specifically, we use the relation of \citet{gh05b}, which was subsequently re-calibrated by \citet{shen11} to be consistent with line+continuum relations (their equation 10). We choose this relation to avoid potential host galaxy contamination of spectra at continuum wavelengths, and also because \ha\ is available for all spectra in the \diskdom\ sample.

We perform spectral fitting on each of the 101 SDSS and INT spectra described in Section S1.3 to recover narrow-line strengths and broad-line strengths and widths to the \ha\ 6563 Å line. The first step of this process uses \gandalf\ \citep{sarzi06} to fit multiple simultaneous lines as well as the continuum. \gandalf\ is optimised for use with SDSS spectra. Using the program with the INT spectra required minimal data re-formatting: we binned logarithmically and de-redshifted the spectra. Initially all emission lines are modeled as a single Gaussian with the same width for all lines. For the SDSS spectra, more information is available from examination of the \oiii\ (4959, 5007) doublet emission-line shape. In 60\% of SDSS sources, a blueshifted wing is required to achieve a satisfactory fit to the \oiii\ narrow-line profile. (The \feii\ emission is simultaneously modeled where necessary.) In 1 source, 3 components are required to model the \oiii\ emission. These findings are consistent with the complex line profiles found by other groups analyzing quasar spectra \citep{shen11}. 

We subsequently focus our analysis on the \ha\ region of the continuum-subtracted spectra, using \emcee\ \citep{emcee13} to simultaneously model the broad \ha , narrow \ha\ and \nii\ doublet emission. We fit the narrow lines using the best-fit profile shape to the \oiii\ line; the \nii\ wavelengths are constrained to a fixed wavelength relative to the narrow \ha\ wavelength and the \nii\ doublet flux ratio is constrained to match the expected ratio between the lines. The broad \ha\ component is modeled with a single component, either a Gaussian or a Lorentzian. We choose the Gaussian model unless the Lorentzian model is clearly a better fit (this is the case for 36 sources). When using the \ha\ FWHM and broad-line luminosity to estimate black hole masses, we use a bootstrap method (with replacement) to sample each value $10^5$ times within the uncertainties, and combine the uncertainties in the resultant black hole masses with those reported by the relation in \citet{gh05b} to arrive at uncertainties for the resulting black hole mass estimates. 

The black hole masses for the sample range from $2 \times 10^6~\mmsun \leq \mmbh \leq 9 \times 10^8~\mmsun$. The median mass is $4 \times 10^7~\mmsun$. We compare these to galaxy stellar masses and bulge mass upper limits in Section \ref{sec:bhmassrelations}. For the 38 sources in the \diskdom\ sample which also have independently estimated black hole masses reported by \citet{shen11}, the masses agree very well, with a median offset of $\sim 0.05$~dex and a dispersion of $\sim 0.2$, similar to what \citeauthor{shen11} measure in their own comparisons between different black hole mass estimators.

%
%
\section{AGN Luminosities, Accretion, and Eddington Ratios}\label{sec:lum}
%
%

\begin{figure*}
\centering{
\includegraphics[width=\textwidth]{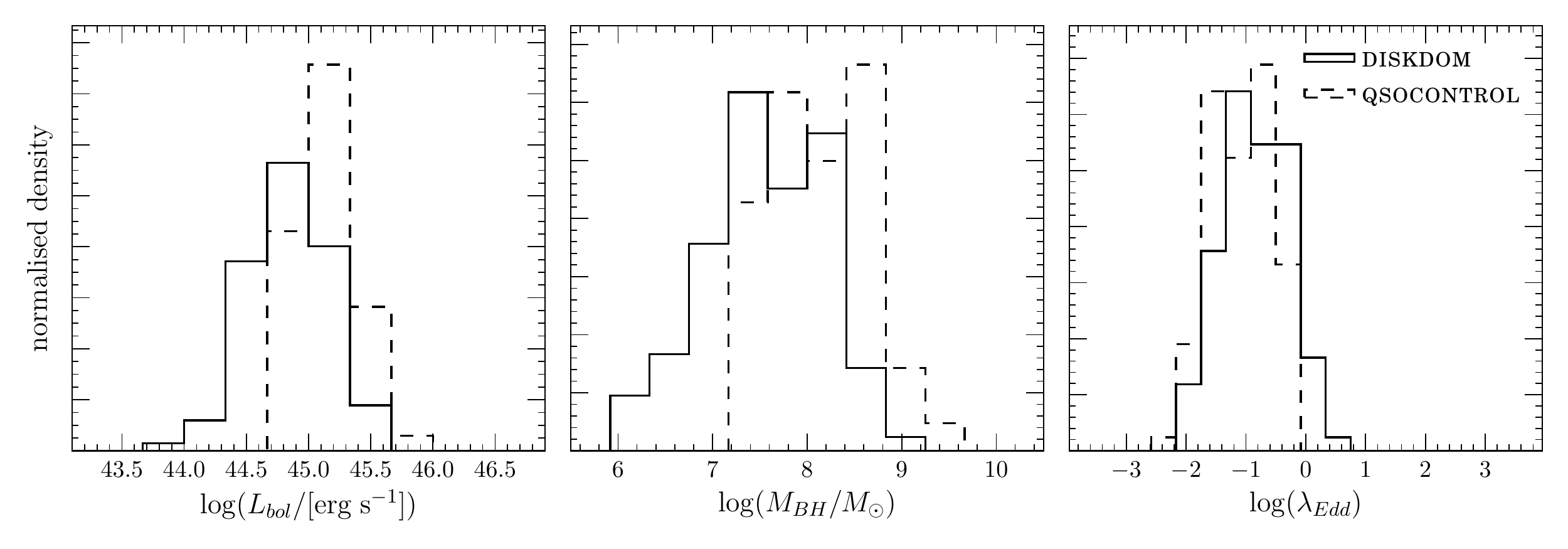}}
\caption{Black hole properties in disk-dominated and control samples: panels show normalised distributions of bolometric luminosity, \lbol\ (left), black hole mass \mbh\ (middle), and \eddrat\ (right) for both the \diskdom\ (solid) and \qsocontrol\ (dashed) samples. For the \lbol\ and \mbh\ distributions, we reject the null hypothesis that the two samples are drawn from the same population, with $p < 10^{-6}$ (Kolmogorov-Smirnov test) in both cases. For the distribution of fractional black hole growth rate \eddrat , the two samples are more similar, and the null hypothesis cannot be confidently ruled out.}
\label{fig:bhproperties}
\end{figure*}

\begin{figure}
\centering{
\includegraphics[width=0.5\textwidth]{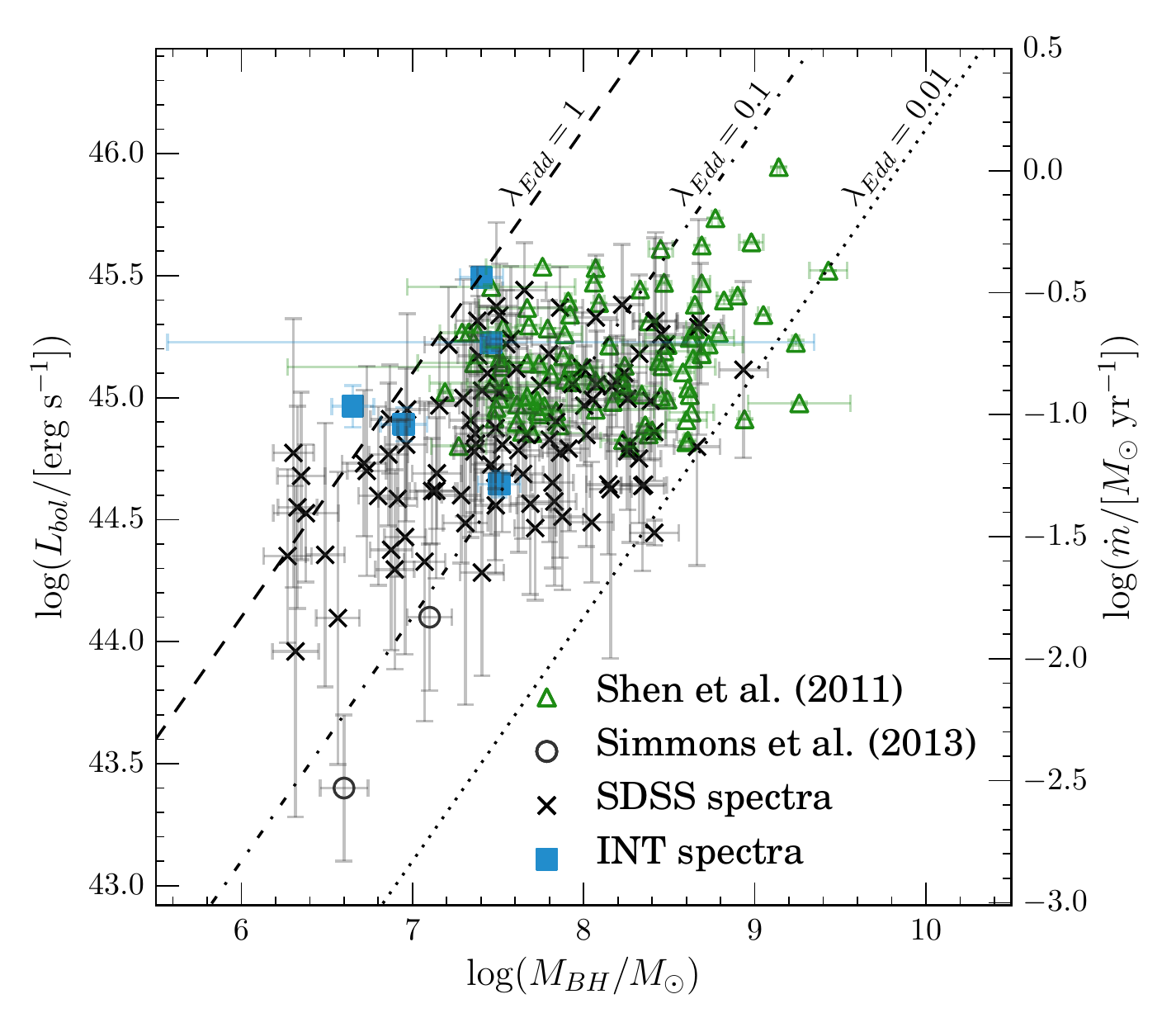}}
\caption{Black hole mass versus bolometric luminosity for the 101 galaxies in the \diskdom\ sample, including those observed by SDSS (black crosses) and with the INT (blue squares). We also show detections from \citet{simmons13} (black open circles) and those from the \qsocontrol\ sample (green triangles). For reference we show lines of example Eddington ratios of $\lambda_{Edd} = 1$ (dashed),  $\lambda_{Edd}$ = 0.1 (dot-dashed) and $\lambda_{Edd} = 0.01$ (dotted). The luminosity range of the \diskdom\ sample includes that of the \qsocontrol\ sample but also extends to lower accretion rates; the black hole masses likewise span a wider range that extends to lower masses.
}
\label{fig:mbhvsbol}
\end{figure}

We estimate bolometric AGN luminosities, \lbol , for the \diskdom\ sample using the \emph{WISE} W3 band at $12 \mu {\rm m}$, applying a correction from \citet{richards06}. The bolometric correction from W3 luminosity is a factor of $\approx 8$, and does not depend significantly on wavelength in the infrared (see their Figure 12). The bolometric luminosities for the \qsocontrol\ sample are provided by \cite{shen11}, who similarly use the bolometric corrections provided by \citet{richards06} and correct from $L_{5100}$ by a factor of $9.26$ for the $z \leq 0.7$ portion of their sample. The difference in choice of wavelength for the bolometric correction does not significantly affect comparisons between samples. The AGN bolometric luminosity is related to its mass accretion rate, \mdot , by a radiative efficiency factor, which we take to be $f = 0.15$ \citep{elvis02} when estimating \mdot .

We additionally use the black hole masses of the samples to compute the luminosities expected if each AGN were accreting at its Eddington limit, \ledd . The Eddington ratio, $\meddrat \equiv \mlbol / \mledd$, is a useful measure of the fractional growth rate of a black hole relative to the maximum rate it can sustain. The bulk of the \diskdom\ sample (75\%) has $0.05 \leq \meddrat \leq 1.0$, with a smaller number of sources having very slow fractional growth ($\meddrat < 0.05$; 20\% of the sample) or super-Eddington accretion ($\meddrat > 1$; 5\% of the sample, though 4 of 5 sources are consistent with Eddington-limited accretion within $1 \sigma$ uncertainties). The average Eddington rate for the sample is $\langle \meddrat \rangle = 0.15$; for the \qsocontrol\ sample it is $\langle \meddrat \rangle = 0.08$. The \qsocontrol\ sample also has a tail of sources accreting at very slow fractional rates, but has no sources with super-Eddington accretion.

Figure \ref{fig:bhproperties} shows the distribution of bolometric AGN luminosity (\lbol ), black hole mass (\mbh ), and Eddington ratio (\eddrat ) for both the \diskdom\ and \qsocontrol\ samples. These properties have overlapping ranges for both samples, but their mass and luminosity distributions are different. The median luminosity of the \diskdom\ sample is lower than that of the \qsocontrol\ sample by a factor of $\sim 2$ ($\mlbol = 7 \times 10^{44}$~erg~s$^{-1}$ for the \diskdom\ sample versus $1 \times 10^{45}$~erg~s$^{-1}$ for the \qsocontrol\ sample), and the most luminous sources in each sample differ by a similar amount, but there is a lower-luminosity tail to the \diskdom\ distribution that reaches a factor of $\sim 6$ lower than that of the \qsocontrol\ sample. 

The black hole mass distributions show a similar pattern: the median mass of the \diskdom\ sample is lower by a factor of $\sim 4$ ($3.5 \times 10^7~\mmsun$ versus $1.4 \times 10^8~\mmsun$ for the \qsocontrol\ sample), with the \diskdom\ sample having a few sources with masses $\sim 9$ times smaller than the smallest black hole in the \qsocontrol\ sample, but the maximum masses ($9 \times 10^8$ and $3 \times 10^9~\mmsun$, respectively) showing a similar offset as the median values. For both \mbh\ and \lbol\ comparisons, the hypothesis that the distributions for the \diskdom\ and \qsocontrol\ samples are drawn from the same population is ruled out to high statistical significance (Kolomogorov-Smirnov $p < 10^{-6}$). Black holes of significant mass are therefore present in the \diskdom\ sample, despite their lack of major merger activity.  

While it is true that the black holes in the \diskdom\ sample span a wide range of masses and accretion rates, as a population they are both less massive and accreting more slowly than the \qsocontrol\ sample. The average \emph{fractional} black hole growth rate $\langle \meddrat \rangle$ is therefore similar in each sample. We cannot rule out the null hypothesis that the two distributions of \eddrat\ are drawn from the same parent distribution. A small number of black holes in the \diskdom\ sample appear to be undergoing super-Eddington accretion, and this sample of galaxies is currently undergoing significant black hole growth.

Figure \ref{fig:mbhvsbol} shows this as a plot of \lbol\ versus \mbh , where lines of constant \eddrat\ are parallel, diagonal lines. The Figure also shows the 2 broad-line AGN from the sample of \citet{simmons13}, which have lower luminosities and black hole masses than the bulk of the \diskdom\ sample. The black hole growth rate, either as a fraction of the maximum rate or in \msun\ yr$^{-1}$, is considerably higher in this sample than in the previous sample of \citeauthor{simmons13}. In other words, the observed accretion rates are high compared to previous studies of AGN in disk-dominated and bulgeless galaxies \citep{filippenko03,satyapal09,arayasalvo12,secrest12,secrest13,simmons13,marleau13,satyapal14,bizzocchi14}. 

At the radiative efficiency value of $f = 0.15$ \citep{elvis02}, we estimate the black hole mass accretion rates to be between $0.01 \leq \mmdot \leq 0.37~\mmsun$~yr$^{-1}$. This \mdot\ range can occur in the absence of significant merger activity: simulations \citep{crockett11,diteodoro14} show that these accretion rates are achievable by cold accretion of minor satellites with masses less than 10\% that of the main galaxy. For some galaxies in the \diskdom\ sample, the images seem to show low surface-brightness satellite galaxies which could indicate growth via ``micro-mergers'' \citep{beaton14}, but careful, dedicated observations will be required to compare the incidence of such galaxies to suitable control samples. 

Using the accretion rates of the \diskdom\ sample, we can estimate the time required to grow a black hole from a seed mass \citep{volonteri08}. We assume the black holes undergo Eddington-limited growth until they reach the observed luminosity, at which point they grow at a constant mass accretion rate (and a decreasing \eddrat ). This is a simple but conservative method to estimate the total time a black hole in this sample must spend actively growing to reach the mass at which it is observed. 

Under these assumptions, the median time spent actively growing for a black hole in the \diskdom\ sample is $\sim 1$~Gyr. This is approximately 7\% of a Hubble time, similar to the AGN duty cycle fraction predicted by independent models and observed in other studies \citep{hopkins06aa,fiore12}. As a population, luminous AGN in disk-dominated galaxies do not appear to be extremes of black hole growth despite their unusual dynamical histories.

We note, however, that for 3 of the most massive black holes in the \diskdom\ sample, the time required to grow from their seed masses to the observed masses is $> 6$ Gyr, indicating that these black holes were likely accreting at a higher rate in the past in order to grow to their current masses given typical duty cycles. These 3 examples as well as the range of observed \mdot\ values support the idea of variable AGN accretion \citep{hopkins09a,schawinski15}. In order to grow the outlier black holes to their current masses within the median time computed for the \diskdom\ sample, the past accretion must have been $\mmdot \sim 3~\mmsun$~yr$^{-1}$ on average. Similar gas inflow rates caused by the presence of a bar have been seen in simulations (\citealt{friedli93}; for additional simulations showing merger-free gas inflow to the inner regions of galaxies see, e.g., \citealt{ciotti91,sakamoto96,maciejewski02a,regan04,hopkins06c,ciotti12,lin13}, and for observational evidence see, e.g., \citealt{davies07,galloway15}). This suggests these growth rates could in theory have been sustained by secular processes.

Regardless of the specific accretion histories of these systems, it is clear that the disk-dominated sample of galaxies we study have grown substantial black holes through a process which is not driven by major mergers. We can now use correlations between the black hole mass and galaxy properties to investigate the nature of this process and its relation to those which drive the assembly of the main galaxy.

%
%
\section{Galaxy-Black Hole Mass Relations}\label{sec:bhmassrelations}
%
%
 
\begin{figure*}
\centering{
\includegraphics[width=0.95\textwidth]{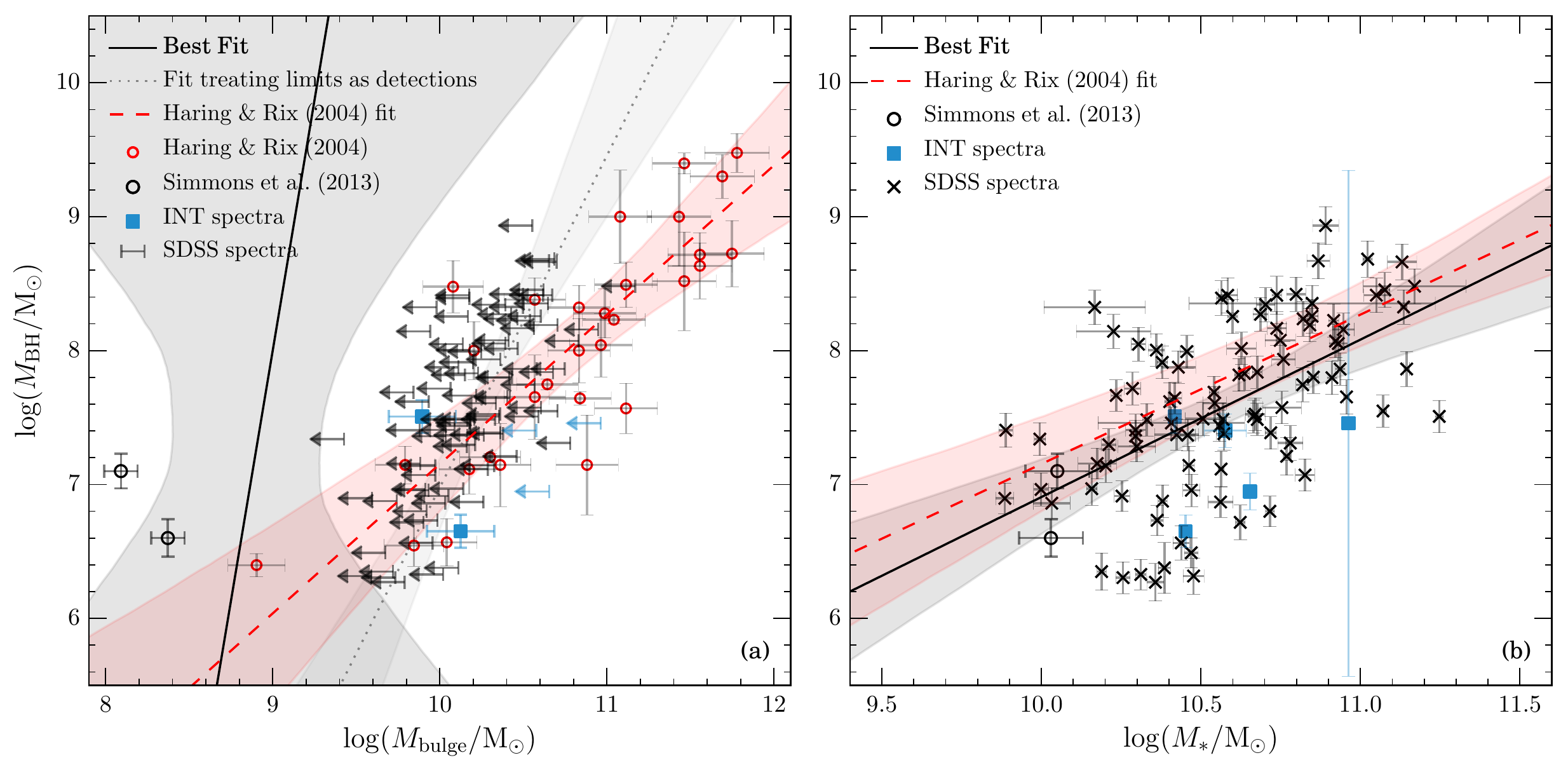}}
\caption{
(a) \textbf{Black hole-bulge relations:} Black hole mass versus bulge stellar mass. Black or blue points show disk-dominated galaxies with bulge masses or upper limits. The best fit line to these data is shown as a solid line, with regions of $\pm 3 \sigma$ uncertainty shown as a shaded region. The fit properly incorporates the bulge mass upper limits as censored data.
The dotted line and $3 \sigma$ shading shows the fit if all upper limits are treated as secure measurements. 
Red open circles show early-type galaxies used to compute a canonical bulge-black hole relation \citep{haringrix04}; we apply the same fitting method to these data and show the best fit line (red-dashed line) and $3 \sigma$ uncertainties (red shaded region).
(b) \textbf{Black hole-galaxy relations:} Black hole mass versus \emph{total} stellar mass. The best-fit line to the \diskdom\ sample is again shown in black (solid line), with the relation for early-types \citep{haringrix04} shown in red (dashed line; unchanged from panel a). The relations are consistent with one another. While the \diskdom\ sample is consistent (in the left panel) with having no correlation between black hole mass and bulge mass and inconsistent with the relation between black hole mass and the bulges of early-type galaxies, the correlations between black hole mass and total stellar mass (right panel) are consistent for populations of disk-dominated and early-type galaxies, despite the very different dynamical and morphological configurations in these samples.}
\label{fig:bhvsgal}
\end{figure*}

We start by comparing black hole mass to the stellar mass attributed to the bulge (see Section \ref{sec:galmass}) for the \diskdom\ sample and for the two unobscured AGN from \citet{simmons13} for which black hole masses were reported. We fit a linear regression model to these sources using a Bayesian method which includes two-dimensional uncertainties \citep[and incorporates upper limits;][]{kelly07}\footnote{\textsc{linmix}: \url{http://linmix.readthedocs.org/}}, and show the results in Figure \ref{fig:bhvsgal}a. We fit both the case where the bulge mass upper limits are treated as real detections, and the case where the upper limits are properly incorporated in a censored fit. When the limits are incorporated into the fit, there is no statistically significant correlation between black hole mass and bulge mass for the disk-dominated sample. 

We follow a similar procedure for a sample of early-type galaxies from \citet{haringrix04}, which have morphologies that indicate an evolutionary history including major mergers \citep{hopkins09a}. The black hole-bulge relation for the early-type galaxies of \citeauthor{haringrix04} is clearly inconsistent with the fitted relation to disk-dominated galaxies (this result persists even in the most conservative case where we re-fit the disk-dominated relation treating all upper limits as real detections; Figure \ref{fig:bhvsgal}a). Black holes do not appear to correlate with bulges in the disk-dominated galaxies of our sample, whereas they do correlate with the masses of bulge-dominated galaxies. 

This result is perhaps unsurprising given that Figure \ref{fig:bhvsgal}a examines bulge-black hole relations in a sample specifically selected to \emph{lack} dominant bulges. Indeed, under the hypothesis that SMBH growth and the black-hole galaxy relation are driven by mergers substantial enough to create bulges one might expect a lack of correlation between black holes and galaxies with merger-free histories. However, we note that the black hole masses in the \diskdom\ sample \emph{have} grown substantially in the absence of significant mergers, even assuming generous seed black hole masses \citep[up to $\sim 10^6~\mmsun$ in the direct collapse scenario;][]{volonteri08,volonteri10}. The most massive black holes in the \diskdom\ sample have grown to masses comparable to the highest masses in the bulge-dominated sample of \citeauthor{haringrix04} ($\sim 10^9~\mmsun$). Furthermore, 60\% of the galaxies in the \diskdom\ sample lie above the correlation for bulge-dominated galaxies (38 of these lie above the $3\sigma$ uncertainty region for that correlation). The black hole masses in our sample are typically considerably \emph{higher} than one would predict based on a bulge-black hole relation, even using generous upper limits to the galaxies' bulge masses.

This analysis assumes that whatever bulges are present have grown exclusively via merger-driven processes. This assumption is questioned by recent observational work \citep{bell17} examining galaxy stellar halos (whose properties trace cumulative merger histories) which indicates that bulges in nearby Milky Way-mass galaxies have higher masses than those expected if mergers are the dominant process driving either pseudo- or classical bulge growth. The stellar masses of the \diskdom\ sample are similar to those in that study, so it is plausible that the results of \citeauthor{bell17} apply here. If so, then in addition to the supermassive black holes in these disk-dominated galaxies being overmassive compared to predictions based on bulges, the bulges themselves are overmassive compared to predictions based on mergers.

Therefore, some process other than major mergers has driven the substantial and ongoing black hole growth in the \diskdom\ sample. As noted in Section \ref{sec:lum}, ``micro-mergers'', disk instabilities, and completely calm ``secular'' processes are all potential causes of SMBH growth. Each of these may produce a different overall rate of black hole growth in galaxies, which may mean galaxies whose growth is dominated by these processes may show very different black hole-galaxy correlations (or none at all) compared to galaxies with growth histories dominated by significant mergers.

It is useful at this point to note that, in determining bulge masses for their bulge-dominated sample, \citet{haringrix04} did not perform bulge-disk decompositions for 29 of the 30 galaxies in their sample. In addition, 26 of those galaxies are classified as either E or S0 types. Thus the $M_{\rm bulge}$ values determined for that study are approximately equivalent (within a small factor) to total stellar masses for $\gtrsim 90$ per cent of that sample. It is therefore appropriate to compare this relation to the fitted relation between black hole mass and total stellar mass for the 101 disk-dominated galaxies in the \diskdom\ sample. Figure \ref{fig:bhvsgal}b shows both relations. 

For both disk-dominated and bulge-dominated galaxies there is a strong correlation between black hole mass and the total stellar mass of the galaxy. The two relations fit to each sample are consistent with each other. We note that the dynamical mass method used by \citeauthor{haringrix04} tends to slightly underestimate stellar masses compared to the colour-based estimates used in Section \ref{sec:galmass} \citep{cappellari06,dejong07}. This systematic difference is consistent with the small difference in the intercept between the two relations in Figure \ref{fig:bhvsgal}b. The slopes of the relations between the early-type sample and the \diskdom\ sample are statistically consistent. The growth of black holes and galaxies in early-type systems with histories including major mergers appears to lead to the same black hole-galaxy co-evolution as that in systems whose dominant disks have been undisturbed by major mergers.

This has significant implications for the origin of the observed black hole-galaxy relations. The merger-driven co-evolution idea \citep{magorrian98,ferrarese00,kormendy01,croton06,hopkins06b,kormendy13} proposes that the correlation between black hole mass and galaxy properties originates with a process which inevitably, fundamentally alters the dynamical and morphological configuration of a galaxy's stars. The observation of the same correlation between black holes and galaxies with morphologies that require an \emph{absence} of major mergers since $z \sim 2$ \citep{martig12} indicates that major galaxy mergers cannot be the sole cause of the correlation. 

It seems clear that both merger-driven and merger-free processes can lead to black hole growth. That the black hole-galaxy correlations are so similar in populations of galaxies with both merger-driven and merger-free histories, however, requires that the growth rates in both galaxies and black holes due to these disparate processes each lead, on average, to the same end point. This is either an impressive coincidence or an indication of a more fundamental cause.  It may be, for example, that AGN feedback \citep[e.g.][]{silk98,fabian99,king03,tortora09} self-regulates both black hole growth and star formation so as to lead to the same black hole and galaxy mass ratio regardless of the process that brings matter to the SMBH. Or, a broader property of galaxies may regulate the amount of material available to any process that causes matter to accrete onto a SMBH.

The gravitational potential would seem to be a plausible candidate for the latter. \citet{vandenbosch16} has recently shown that the virial theorem implies black hole masses primarily correlate with global velocity dispersions rather than bulge properties, for any galaxy (of any type) that hosts a supermassive black hole. Previous theoretical studies find that black hole masses correlate with galaxy baryonic properties insofar as both reflect the overall gravitational potential \citep{booth10,volonteri11}, i.e., that the observed black hole-galaxy connection is an expected outcome of hierarchical galaxy evolution \citep{jahnke11}, which includes merger-free growth via cold accretion \citep{keres05,welker15}. Our results would seem to support this hypothesis; further observations to characterise the velocity dispersions and extended disk rotational velocities for the \diskdom\ sample could provide a test of the predictions of \citet{vandenbosch16}.

%
%
\section{Summary}\label{sec:summary}
%
%

We have identified a sample of 101 unobscured AGN with bolometric luminosities from $9 \times 10^{43} \lesssim \mlbol \lesssim 3 \times 10^{45}$~erg~s$^{-1}$ and hosted in unambiguously disk-dominated galaxies. The luminosities and accretion rates of the black holes reach into the luminosity regime of quasars at similar redshifts ($z \leq 0.25$) despite being hosted in galaxies whose morphologies indicate evolutionary histories since $z \sim 2$ free from significant mergers that would have created substantial bulges. 

Black hole masses for the sample, estimated via broadened \ha\ line emission in single-epoch SDSS and INT spectra, are between $2 \times 10^6~\mmsun \leq \mmbh \leq 9 \times 10^8~\mmsun$, with a median mass of $4 \times 10^7~\mmsun$. The black hole masses are typically slightly smaller ($0.6$ dex median difference) than a redshift-matched comparison sample of SDSS quasars. However, their fractional growth rates are similar, with median $\meddrat \approx 0.2$ for the disk-dominated sample versus $0.1$ for the comparison quasar sample, whose morphologies are more consistent with early-type galaxies.

Comparing black hole masses with upper limits to bulge masses for the disk-dominated sample, we find that a fitted relation between the two quantities is consistent with an undefined slope, i.e., no correlation. Moreover, it is statistically inconsistent with the canonical bulge-black hole relation of \citet{haringrix04}, which we have re-fitted to their sample using the same Bayesian method as for the disk-dominated relation.

As for the relationship between black hole mass and total stellar mass, we find a fitted relation for the disk-dominated quasar sample that agrees very well with that of \citet{haringrix04}, which is based on bulge-dominated and early-type galaxies. That black holes and galaxies appear to reach the same correlation in both disk-dominated and bulge-dominated galaxies implies that the different evolutionary processes (such as major mergers versus secular processes) leading to these two dynamically and morphologically distinct populations are not fundamental drivers of black hole-galaxy co-evolution.

%
%
\section*{Acknowledgments}
%
%
{\changed The authors wish to thank M. Cappellari, A. Coil, E. Glikman, E. Moran, and C. M. Urry for many useful discussions related to this work. The authors are also grateful to the anonymous referee for suggestions which improved this manuscript.}

{\changed In addition to the software cited in the main text,} this research made extensive use of the the Python package {\tt astroPy}\footnote{\url{http://www.astropy.org/}} \citep{astropy13}, NASA's ADS service and Cornell's ArXiv. This publication also used an OS X widget form of the JavaScript Cosmology Calculator \citep{wright06,rsimpson13} and the Tool for Operations on Catalogues And Tables (TOPCAT; ~\citealt{Taylor05})\footnote{\url{http://www.star.bris.ac.uk/~mbt/topcat/}}.

BDS gratefully acknowledges support from Balliol College, Oxford through the Henry Skynner Junior Research Fellowship, and from the National Aeronautics and Space Administration (NASA) through Einstein Postdoctoral Fellowship Award Number PF5-160143 issued by the Chandra X-ray Observatory Center, which is operated by the Smithsonian Astrophysical Observatory for and on behalf of NASA under contract NAS8-03060. 
RJS acknowledges funding from the Science and Technology Facilities Council Grant Code ST/K502236/1 and from the Ogden Trust.

Funding for the SDSS and SDSS-II has been provided by the Alfred P. Sloan Foundation, the Participating Institutions, the National Science Foundation, the U.S. Department of Energy, the National Aeronautics and Space Administration, the Japanese Monbukagakusho, the Max Planck Society, and the Higher Education Funding Council for England. The SDSS Web Site is \url{http://www.sdss.org/}.
The SDSS is managed by the Astrophysical Research Consortium for the Participating Institutions. The Participating Institutions are the American Museum of Natural History, Astrophysical Institute Potsdam, University of Basel, University of Cambridge, Case Western Reserve University, University of Chicago, Drexel University, Fermilab, the Institute for Advanced Study, the Japan Participation Group, Johns Hopkins University, the Joint Institute for Nuclear Astrophysics, the Kavli Institute for Particle Astrophysics and Cosmology, the Korean Scientist Group, the Chinese Academy of Sciences (LAMOST), Los Alamos National Laboratory, the Max-Planck-Institute for Astronomy (MPIA), the Max-Planck-Institute for Astrophysics (MPA), New Mexico State University, Ohio State University, University of Pittsburgh, University of Portsmouth, Princeton University, the United States Naval Observatory, and the University of Washington.

This publication makes use of data products from the Wide-field Infrared Survey Explorer, which is a joint project of the University of California, Los Angeles, and the Jet Propulsion Laboratory/California Institute of Technology, funded by the National Aeronautics and Space Administration.
\bibliographystyle{mnras}
\bibliography{refs}  

\begin{thebibliography}{}
\makeatletter
\relax
\def\mn@urlcharsother{\let\do\@makeother \do\$\do\&\do\#\do\^\do\_\do\%\do\~}
\def\mn@doi{\begingroup\mn@urlcharsother \@ifnextchar [ {\mn@doi@}
  {\mn@doi@[]}}
\def\mn@doi@[#1]#2{\def\@tempa{#1}\ifx\@tempa\@empty \href
  {http://dx.doi.org/#2} {doi:#2}\else \href {http://dx.doi.org/#2} {#1}\fi
  \endgroup}
\def\mn@eprint#1#2{\mn@eprint@#1:#2::\@nil}
\def\mn@eprint@arXiv#1{\href {http://arxiv.org/abs/#1} {{\tt arXiv:#1}}}
\def\mn@eprint@dblp#1{\href {http://dblp.uni-trier.de/rec/bibtex/#1.xml}
  {dblp:#1}}
\def\mn@eprint@#1:#2:#3:#4\@nil{\def\@tempa {#1}\def\@tempb {#2}\def\@tempc
  {#3}\ifx \@tempc \@empty \let \@tempc \@tempb \let \@tempb \@tempa \fi \ifx
  \@tempb \@empty \def\@tempb {arXiv}\fi \@ifundefined
  {mn@eprint@\@tempb}{\@tempb:\@tempc}{\expandafter \expandafter \csname
  mn@eprint@\@tempb\endcsname \expandafter{\@tempc}}}

\bibitem[\protect\citeauthoryear{{Abazajian} et~al.,}{{Abazajian}
  et~al.}{2009}]{abazajian09}
{Abazajian} K.~N.,  et~al., 2009, \mn@doi [\apjs]
  {10.1088/0067-0049/182/2/543}, \href
  {http://adsabs.harvard.edu/abs/2009ApJS..182..543A} {182, 543}

\bibitem[\protect\citeauthoryear{{Ahn} et~al.,}{{Ahn} et~al.}{2012}]{ahn12}
{Ahn} C.~P.,  et~al., 2012, \mn@doi [\apjs] {10.1088/0067-0049/203/2/21}, \href
  {http://adsabs.harvard.edu/abs/2012ApJS..203...21A} {203, 21}

\bibitem[\protect\citeauthoryear{{Aihara} et~al.,}{{Aihara}
  et~al.}{2011}]{aihara11}
{Aihara} H.,  et~al., 2011, \mn@doi [\apjs] {10.1088/0067-0049/193/2/29}, \href
  {http://adsabs.harvard.edu/abs/2011ApJS..193...29A} {193, 29}

\bibitem[\protect\citeauthoryear{{Araya Salvo}, {Mathur}, {Ghosh}, {Fiore}  \&
  {Ferrarese}}{{Araya Salvo} et~al.}{2012}]{arayasalvo12}
{Araya Salvo} C.,  {Mathur} S.,  {Ghosh} H.,  {Fiore} F.,   {Ferrarese} L.,
  2012, \mn@doi [\apj] {10.1088/0004-637X/757/2/179}, \href
  {http://adsabs.harvard.edu/abs/2012ApJ...757..179A} {757, 179}

\bibitem[\protect\citeauthoryear{{Astropy Collaboration} et~al.,}{{Astropy
  Collaboration} et~al.}{2013}]{astropy13}
{Astropy Collaboration} et~al., 2013, \mn@doi [\aap]
  {10.1051/0004-6361/201322068}, \href
  {http://adsabs.harvard.edu/abs/2013A%26A...558A..33A} {558, A33}

\bibitem[\protect\citeauthoryear{{Baldry}, {Balogh}, {Bower}, {Glazebrook},
  {Nichol}, {Bamford}  \& {Budavari}}{{Baldry} et~al.}{2006}]{baldry06}
{Baldry} I.~K.,  {Balogh} M.~L.,  {Bower} R.~G.,  {Glazebrook} K.,  {Nichol}
  R.~C.,  {Bamford} S.~P.,   {Budavari} T.,  2006, \mn@doi [\mnras]
  {10.1111/j.1365-2966.2006.11081.x}, \href
  {http://adsabs.harvard.edu/abs/2006MNRAS.373..469B} {373, 469}

\bibitem[\protect\citeauthoryear{{Bauer}, {Alexander}, {Brandt}, {Schneider},
  {Treister}, {Hornschemeier}  \& {Garmire}}{{Bauer} et~al.}{2004}]{bauer04}
{Bauer} F.~E.,  {Alexander} D.~M.,  {Brandt} W.~N.,  {Schneider} D.~P.,
  {Treister} E.,  {Hornschemeier} A.~E.,   {Garmire} G.~P.,  2004, \mn@doi
  [\aj] {10.1086/424859}, \href
  {http://adsabs.harvard.edu/abs/2004AJ....128.2048B} {128, 2048}

\bibitem[\protect\citeauthoryear{{Beaton}}{{Beaton}}{2014}]{beaton14}
{Beaton} R.~L.,  2014, PhD thesis, University of Virginia

\bibitem[\protect\citeauthoryear{{Bell} \& {de Jong}}{{Bell} \& {de
  Jong}}{2001}]{bell01}
{Bell} E.~F.,  {de Jong} R.~S.,  2001, \mn@doi [\apj] {10.1086/319728}, \href
  {http://adsabs.harvard.edu/abs/2001ApJ...550..212B} {550, 212}

\bibitem[\protect\citeauthoryear{{Bell}, {Monachesi}, {Harmsen}, {de Jong},
  {Bailin}, {Radburn-Smith}, {D'Souza}  \& {Holwerda}}{{Bell}
  et~al.}{2017}]{bell17}
{Bell} E.~F.,  {Monachesi} A.,  {Harmsen} B.,  {de Jong} R.~S.,  {Bailin} J.,
  {Radburn-Smith} D.~J.,  {D'Souza} R.,   {Holwerda} B.~W.,  2017, \mn@doi
  [\apjl] {10.3847/2041-8213/aa6158}, \href
  {http://adsabs.harvard.edu/abs/2017ApJ...837L...8B} {837, L8}

\bibitem[\protect\citeauthoryear{{Bennett} et~al.,}{{Bennett}
  et~al.}{2013}]{bennett13}
{Bennett} C.~L.,  et~al., 2013, \mn@doi [\apjs] {10.1088/0067-0049/208/2/20},
  \href {http://adsabs.harvard.edu/abs/2013ApJS..208...20B} {208, 20}

\bibitem[\protect\citeauthoryear{{Bessiere}, {Tadhunter}, {Ramos Almeida}  \&
  {Villar Mart{\'{\i}}n}}{{Bessiere} et~al.}{2014}]{bessiere14}
{Bessiere} P.~S.,  {Tadhunter} C.~N.,  {Ramos Almeida} C.,   {Villar
  Mart{\'{\i}}n} M.,  2014, \mn@doi [\mnras] {10.1093/mnras/stt2333}, \href
  {http://adsabs.harvard.edu/abs/2014MNRAS.438.1839B} {438, 1839}

\bibitem[\protect\citeauthoryear{{Bizzocchi} et~al.,}{{Bizzocchi}
  et~al.}{2014}]{bizzocchi14}
{Bizzocchi} L.,  et~al., 2014, \mn@doi [\apj] {10.1088/0004-637X/782/1/22},
  \href {http://adsabs.harvard.edu/abs/2014ApJ...782...22B} {782, 22}

\bibitem[\protect\citeauthoryear{{Booth} \& {Schaye}}{{Booth} \&
  {Schaye}}{2010}]{booth10}
{Booth} C.~M.,  {Schaye} J.,  2010, \mn@doi [\mnras]
  {10.1111/j.1745-3933.2010.00832.x}, \href
  {http://adsabs.harvard.edu/abs/2010MNRAS.405L...1B} {405, L1}

\bibitem[\protect\citeauthoryear{{Bournaud} et~al.,}{{Bournaud}
  et~al.}{2012}]{bournaud12}
{Bournaud} F.,  et~al., 2012, \mn@doi [\apj] {10.1088/0004-637X/757/1/81},
  \href {http://adsabs.harvard.edu/abs/2012ApJ...757...81B} {757, 81}

\bibitem[\protect\citeauthoryear{{Brook}, {Stinson}, {Gibson}, {Ro{\v s}kar},
  {Wadsley}  \& {Quinn}}{{Brook} et~al.}{2012}]{brook12}
{Brook} C.~B.,  {Stinson} G.,  {Gibson} B.~K.,  {Ro{\v s}kar} R.,  {Wadsley}
  J.,   {Quinn} T.,  2012, \mn@doi [\mnras] {10.1111/j.1365-2966.2011.19740.x},
  \href {http://adsabs.harvard.edu/abs/2012MNRAS.419..771B} {419, 771}

\bibitem[\protect\citeauthoryear{{Cappellari} et~al.,}{{Cappellari}
  et~al.}{2006}]{cappellari06}
{Cappellari} M.,  et~al., 2006, \mn@doi [\mnras]
  {10.1111/j.1365-2966.2005.09981.x}, \href
  {http://adsabs.harvard.edu/abs/2006MNRAS.366.1126C} {366, 1126}

\bibitem[\protect\citeauthoryear{{Ciotti} \& {Ostriker}}{{Ciotti} \&
  {Ostriker}}{2012}]{ciotti12}
{Ciotti} L.,  {Ostriker} J.~P.,  2012, in {D.-W.~Kim \& S.~Pellegrini} ed.,
  Astrophysics and Space Science Library Vol. 378, Astrophysics and Space
  Science Library. p.~83 (\mn@eprint {arXiv} {1104.2238}),
  \mn@doi{10.1007/978-1-4614-0580-1_4}

\bibitem[\protect\citeauthoryear{{Ciotti}, {D'Ercole}, {Pellegrini}  \&
  {Renzini}}{{Ciotti} et~al.}{1991}]{ciotti91}
{Ciotti} L.,  {D'Ercole} A.,  {Pellegrini} S.,   {Renzini} A.,  1991, \mn@doi
  [\apj] {10.1086/170289}, \href
  {http://adsabs.harvard.edu/abs/1991ApJ...376..380C} {376, 380}

\bibitem[\protect\citeauthoryear{{Cisternas} et~al.,}{{Cisternas}
  et~al.}{2011}]{cisternas11}
{Cisternas} M.,  et~al., 2011, \mn@doi [\apjl] {10.1088/2041-8205/741/1/L11},
  \href {http://adsabs.harvard.edu/abs/2011ApJ...741L..11C} {741, L11}

\bibitem[\protect\citeauthoryear{{Crockett} et~al.,}{{Crockett}
  et~al.}{2011}]{crockett11}
{Crockett} R.~M.,  et~al., 2011, \mn@doi [\apj] {10.1088/0004-637X/727/2/115},
  \href {http://adsabs.harvard.edu/abs/2011ApJ...727..115C} {727, 115}

\bibitem[\protect\citeauthoryear{{Croton} et~al.,}{{Croton}
  et~al.}{2006}]{croton06}
{Croton} D.~J.,  et~al., 2006, \mn@doi [\mnras]
  {10.1111/j.1365-2966.2005.09675.x}, \href
  {http://adsabs.harvard.edu/abs/2006MNRAS.365...11C} {365, 11}

\bibitem[\protect\citeauthoryear{{Davies}, {M{\"u}ller S{\'a}nchez}, {Genzel},
  {Tacconi}, {Hicks}, {Friedrich}  \& {Sternberg}}{{Davies}
  et~al.}{2007}]{davies07}
{Davies} R.~I.,  {M{\"u}ller S{\'a}nchez} F.,  {Genzel} R.,  {Tacconi} L.~J.,
  {Hicks} E.~K.~S.,  {Friedrich} S.,   {Sternberg} A.,  2007, \mn@doi [\apj]
  {10.1086/523032}, \href {http://adsabs.harvard.edu/abs/2007ApJ...671.1388D}
  {671, 1388}

\bibitem[\protect\citeauthoryear{{Di Teodoro} \& {Fraternali}}{{Di Teodoro} \&
  {Fraternali}}{2014}]{diteodoro14}
{Di Teodoro} E.~M.,  {Fraternali} F.,  2014, \mn@doi [\aap]
  {10.1051/0004-6361/201423596}, \href
  {http://adsabs.harvard.edu/abs/2014A%26A...567A..68D} {567, A68}

\bibitem[\protect\citeauthoryear{{Dunlop}, {McLure}, {Kukula}, {Baum}, {O'Dea}
  \& {Hughes}}{{Dunlop} et~al.}{2003}]{dunlop03}
{Dunlop} J.~S.,  {McLure} R.~J.,  {Kukula} M.~J.,  {Baum} S.~A.,  {O'Dea}
  C.~P.,   {Hughes} D.~H.,  2003, \mn@doi [\mnras]
  {10.1046/j.1365-8711.2003.06333.x}, \href
  {http://adsabs.harvard.edu/cgi-bin/nph-bib_query?bibcode=2003MNRAS.340.1095D&db_key=AST}
  {340, 1095}

\bibitem[\protect\citeauthoryear{{Edelson} \& {Malkan}}{{Edelson} \&
  {Malkan}}{2012}]{edelson12}
{Edelson} R.,  {Malkan} M.,  2012, \mn@doi [\apj] {10.1088/0004-637X/751/1/52},
  \href {http://adsabs.harvard.edu/abs/2012ApJ...751...52E} {751, 52}

\bibitem[\protect\citeauthoryear{{Ellison}, {Mendel}, {Patton}  \&
  {Scudder}}{{Ellison} et~al.}{2013}]{ellison13}
{Ellison} S.~L.,  {Mendel} J.~T.,  {Patton} D.~R.,   {Scudder} J.~M.,  2013,
  \mn@doi [\mnras] {10.1093/mnras/stt1562}, \href
  {http://adsabs.harvard.edu/abs/2013MNRAS.435.3627E} {435, 3627}

\bibitem[\protect\citeauthoryear{{Elvis}, {Risaliti}  \& {Zamorani}}{{Elvis}
  et~al.}{2002}]{elvis02}
{Elvis} M.,  {Risaliti} G.,   {Zamorani} G.,  2002, \mn@doi [\apjl]
  {10.1086/339197}, \href {http://adsabs.harvard.edu/abs/2002ApJ...565L..75E}
  {565, L75}

\bibitem[\protect\citeauthoryear{{Emsellem} et~al.,}{{Emsellem}
  et~al.}{2011}]{emsellem11}
{Emsellem} E.,  et~al., 2011, \mn@doi [\mnras]
  {10.1111/j.1365-2966.2011.18496.x}, \href
  {http://adsabs.harvard.edu/abs/2011MNRAS.414..888E} {414, 888}

\bibitem[\protect\citeauthoryear{{Fabian}}{{Fabian}}{1999}]{fabian99}
{Fabian} A.~C.,  1999, \mn@doi [\mnras] {10.1046/j.1365-8711.1999.03017.x},
  \href {http://adsabs.harvard.edu/abs/1999MNRAS.308L..39F} {308, L39}

\bibitem[\protect\citeauthoryear{{Ferrarese} \& {Merritt}}{{Ferrarese} \&
  {Merritt}}{2000}]{ferrarese00}
{Ferrarese} L.,  {Merritt} D.,  2000, \mn@doi [\apjl] {10.1086/312838}, \href
  {http://adsabs.harvard.edu/cgi-bin/nph-bib_query?bibcode=2000ApJ...539L...9F&db_key=AST}
  {539, L9}

\bibitem[\protect\citeauthoryear{{Filippenko} \& {Ho}}{{Filippenko} \&
  {Ho}}{2003}]{filippenko03}
{Filippenko} A.~V.,  {Ho} L.~C.,  2003, \mn@doi [\apjl] {10.1086/375361}, \href
  {http://adsabs.harvard.edu/abs/2003ApJ...588L..13F} {588, L13}

\bibitem[\protect\citeauthoryear{{Fiore} et~al.,}{{Fiore}
  et~al.}{2012}]{fiore12}
{Fiore} F.,  et~al., 2012, \mn@doi [\aap] {10.1051/0004-6361/201117581}, \href
  {http://adsabs.harvard.edu/abs/2012A%26A...537A..16F} {537, A16}

\bibitem[\protect\citeauthoryear{{Foreman-Mackey}, {Hogg}, {Lang}  \&
  {Goodman}}{{Foreman-Mackey} et~al.}{2013}]{emcee13}
{Foreman-Mackey} D.,  {Hogg} D.~W.,  {Lang} D.,   {Goodman} J.,  2013, \mn@doi
  [\pasp] {10.1086/670067}, \href
  {http://adsabs.harvard.edu/abs/2013PASP..125..306F} {125, 306}

\bibitem[\protect\citeauthoryear{{Friedli} \& {Benz}}{{Friedli} \&
  {Benz}}{1993}]{friedli93}
{Friedli} D.,  {Benz} W.,  1993, \aap, \href
  {http://adsabs.harvard.edu/abs/1993A\%26A...268...65F} {268, 65–85}

\bibitem[\protect\citeauthoryear{{Gabor} et~al.,}{{Gabor}
  et~al.}{2009}]{gabor09}
{Gabor} J.~M.,  et~al., 2009, \mn@doi [\apj] {10.1088/0004-637X/691/1/705},
  \href {http://adsabs.harvard.edu/abs/2009ApJ...691..705G} {691, 705}

\bibitem[\protect\citeauthoryear{{Galloway} et~al.,}{{Galloway}
  et~al.}{2015}]{galloway15}
{Galloway} M.~A.,  et~al., 2015, \mn@doi [\mnras] {10.1093/mnras/stv235}, \href
  {http://adsabs.harvard.edu/abs/2015MNRAS.448.3442G} {448, 3442}

\bibitem[\protect\citeauthoryear{{Glikman}, {Simmons}, {Mailly}, {Schawinski},
  {Urry}  \& {Lacy}}{{Glikman} et~al.}{2015}]{glikman15}
{Glikman} E.,  {Simmons} B.,  {Mailly} M.,  {Schawinski} K.,  {Urry} C.~M.,
  {Lacy} M.,  2015, \mn@doi [\apj] {10.1088/0004-637X/806/2/218}, \href
  {http://adsabs.harvard.edu/abs/2015ApJ...806..218G} {806, 218}

\bibitem[\protect\citeauthoryear{{Graham}}{{Graham}}{2001}]{graham01a}
{Graham} A.~W.,  2001, \mn@doi [\aj] {10.1086/318767}, \href
  {http://adsabs.harvard.edu/abs/2001AJ....121..820G} {121, 820}

\bibitem[\protect\citeauthoryear{{Greene} \& {Ho}}{{Greene} \&
  {Ho}}{2005}]{gh05b}
{Greene} J.~E.,  {Ho} L.~C.,  2005, \mn@doi [\apj] {10.1086/431897}, \href
  {http://adsabs.harvard.edu/abs/2005ApJ...630..122G} {630, 122}

\bibitem[\protect\citeauthoryear{{Greene} \& {Ho}}{{Greene} \&
  {Ho}}{2006}]{gh06}
{Greene} J.~E.,  {Ho} L.~C.,  2006, \mn@doi [\apjl] {10.1086/500507}, \href
  {http://adsabs.harvard.edu/abs/2006ApJ...641L..21G} {641, L21}

\bibitem[\protect\citeauthoryear{{Greene} \& {Ho}}{{Greene} \&
  {Ho}}{2007}]{gh07a}
{Greene} J.~E.,  {Ho} L.~C.,  2007, \mn@doi [\apj] {10.1086/522082}, \href
  {http://adsabs.harvard.edu/abs/2007ApJ...670...92G} {670, 92}

\bibitem[\protect\citeauthoryear{{Greene} et~al.,}{{Greene}
  et~al.}{2010}]{greene10b}
{Greene} J.~E.,  et~al., 2010, \mn@doi [\apj] {10.1088/0004-637X/721/1/26},
  \href {http://adsabs.harvard.edu/abs/2010ApJ...721...26G} {721, 26}

\bibitem[\protect\citeauthoryear{{Grogin} et~al.,}{{Grogin}
  et~al.}{2005}]{grogin05}
{Grogin} N.~A.,  et~al., 2005, \mn@doi [\apjl] {10.1086/432256}, \href
  {http://adsabs.harvard.edu/cgi-bin/nph-bib_query?bibcode=2005ApJ...627L..97G&db_key=AST}
  {627, L97}

\bibitem[\protect\citeauthoryear{{Hao}, {Jogee}, {Barazza}, {Marinova}  \&
  {Shen}}{{Hao} et~al.}{2009}]{hao09}
{Hao} L.,  {Jogee} S.,  {Barazza} F.~D.,  {Marinova} I.,   {Shen} J.,  2009, in
  {Jogee} S.,  {Marinova} I.,  {Hao} L.,   {Blanc} G.~A.,  eds,  Astronomical
  Society of the Pacific Conference Series Vol. 419, Galaxy Evolution: Emerging
  Insights and Future Challenges. p.~402 (\mn@eprint {arXiv} {0910.3960})

\bibitem[\protect\citeauthoryear{{H{\"a}ring} \& {Rix}}{{H{\"a}ring} \&
  {Rix}}{2004}]{haringrix04}
{H{\"a}ring} N.,  {Rix} H.-W.,  2004, \mn@doi [\apjl] {10.1086/383567}, \href
  {http://adsabs.harvard.edu/abs/2004ApJ...604L..89H} {604, L89}

\bibitem[\protect\citeauthoryear{{Hasinger}, {Miyaji}  \& {Schmidt}}{{Hasinger}
  et~al.}{2005}]{hasinger05}
{Hasinger} G.,  {Miyaji} T.,   {Schmidt} M.,  2005, \mn@doi [\aap]
  {10.1051/0004-6361:20042134}, \href
  {http://adsabs.harvard.edu/abs/2005A%26A...441..417H} {441, 417}

\bibitem[\protect\citeauthoryear{{H{\"a}ussler} et~al.,}{{H{\"a}ussler}
  et~al.}{2007}]{haussler07}
{H{\"a}ussler} B.,  et~al., 2007, \mn@doi [\apjs] {10.1086/518836}, \href
  {http://adsabs.harvard.edu/abs/2007ApJS..172..615H} {172, 615}

\bibitem[\protect\citeauthoryear{{Hopkins} \& {Hernquist}}{{Hopkins} \&
  {Hernquist}}{2006}]{hopkins06c}
{Hopkins} P.~F.,  {Hernquist} L.,  2006, \mn@doi [\apjs] {10.1086/505753},
  \href {http://adsabs.harvard.edu/abs/2006ApJS..166....1H} {166, 1}

\bibitem[\protect\citeauthoryear{{Hopkins} \& {Hernquist}}{{Hopkins} \&
  {Hernquist}}{2009}]{hopkins09a}
{Hopkins} P.~F.,  {Hernquist} L.,  2009, \mn@doi [\apj]
  {10.1088/0004-637X/694/1/599}, \href
  {http://adsabs.harvard.edu/abs/2009ApJ...694..599H} {694, 599}

\bibitem[\protect\citeauthoryear{{Hopkins}, {Hernquist}, {Cox}, {Di Matteo},
  {Robertson}  \& {Springel}}{{Hopkins} et~al.}{2006a}]{hopkins06aa}
{Hopkins} P.~F.,  {Hernquist} L.,  {Cox} T.~J.,  {Di Matteo} T.,  {Robertson}
  B.,   {Springel} V.,  2006a, \mn@doi [\apjs] {10.1086/499298}, \href
  {http://adsabs.harvard.edu/abs/2006ApJS..163....1H} {163, 1}

\bibitem[\protect\citeauthoryear{{Hopkins}, {Somerville}, {Hernquist}, {Cox},
  {Robertson}  \& {Li}}{{Hopkins} et~al.}{2006b}]{hopkins06b}
{Hopkins} P.~F.,  {Somerville} R.~S.,  {Hernquist} L.,  {Cox} T.~J.,
  {Robertson} B.,   {Li} Y.,  2006b, \mn@doi [\apj] {10.1086/508503}, \href
  {http://adsabs.harvard.edu/abs/2006ApJ...652..864H} {652, 864}

\bibitem[\protect\citeauthoryear{{Hopkins}, {Cox}, {Younger}  \&
  {Hernquist}}{{Hopkins} et~al.}{2009}]{hopkins09c}
{Hopkins} P.~F.,  {Cox} T.~J.,  {Younger} J.~D.,   {Hernquist} L.,  2009,
  \mn@doi [\apj] {10.1088/0004-637X/691/2/1168}, \href
  {http://adsabs.harvard.edu/abs/2009ApJ...691.1168H} {691, 1168}

\bibitem[\protect\citeauthoryear{{Hopkins}, {Kere{\v s}}, {Murray}, {Quataert}
  \& {Hernquist}}{{Hopkins} et~al.}{2012}]{hopkins11c}
{Hopkins} P.~F.,  {Kere{\v s}} D.,  {Murray} N.,  {Quataert} E.,   {Hernquist}
  L.,  2012, \mn@doi [\mnras] {10.1111/j.1365-2966.2012.21981.x}, \href
  {http://adsabs.harvard.edu/abs/2012MNRAS.427..968H} {427, 968}

\bibitem[\protect\citeauthoryear{{Jahnke} \& {Macci{\`o}}}{{Jahnke} \&
  {Macci{\`o}}}{2011}]{jahnke11}
{Jahnke} K.,  {Macci{\`o}} A.~V.,  2011, \mn@doi [\apj]
  {10.1088/0004-637X/734/2/92}, \href
  {http://adsabs.harvard.edu/abs/2011ApJ...734...92J} {734, 92}

\bibitem[\protect\citeauthoryear{{Jiang}, {Greene}  \& {Ho}}{{Jiang}
  et~al.}{2011a}]{jiang11a}
{Jiang} Y.-F.,  {Greene} J.~E.,   {Ho} L.~C.,  2011a, \mn@doi [\apjl]
  {10.1088/2041-8205/737/2/L45}, \href
  {http://adsabs.harvard.edu/abs/2011ApJ...737L..45J} {737, L45}

\bibitem[\protect\citeauthoryear{{Jiang}, {Greene}, {Ho}, {Xiao}  \&
  {Barth}}{{Jiang} et~al.}{2011b}]{jiang11b}
{Jiang} Y.-F.,  {Greene} J.~E.,  {Ho} L.~C.,  {Xiao} T.,   {Barth} A.~J.,
  2011b, \mn@doi [\apj] {10.1088/0004-637X/742/2/68}, \href
  {http://adsabs.harvard.edu/abs/2011ApJ...742...68J} {742, 68}

\bibitem[\protect\citeauthoryear{{Kannan}, {Macci{\`o}}, {Fontanot}, {Moster},
  {Karman}  \& {Somerville}}{{Kannan} et~al.}{2015}]{kannan15}
{Kannan} R.,  {Macci{\`o}} A.~V.,  {Fontanot} F.,  {Moster} B.~P.,  {Karman}
  W.,   {Somerville} R.~S.,  2015, \mn@doi [\mnras] {10.1093/mnras/stv1633},
  \href {http://adsabs.harvard.edu/abs/2015MNRAS.452.4347K} {452, 4347}

\bibitem[\protect\citeauthoryear{{Kelly}}{{Kelly}}{2007}]{kelly07}
{Kelly} B.~C.,  2007, \mn@doi [\apj] {10.1086/519947}, \href
  {http://adsabs.harvard.edu/abs/2007ApJ...665.1489K} {665, 1489}

\bibitem[\protect\citeauthoryear{{Kere{\v s}}, {Katz}, {Weinberg}  \&
  {Dav{\'e}}}{{Kere{\v s}} et~al.}{2005}]{keres05}
{Kere{\v s}} D.,  {Katz} N.,  {Weinberg} D.~H.,   {Dav{\'e}} R.,  2005, \mn@doi
  [\mnras] {10.1111/j.1365-2966.2005.09451.x}, \href
  {http://adsabs.harvard.edu/abs/2005MNRAS.363....2K} {363, 2}

\bibitem[\protect\citeauthoryear{{Kim}, {Ho}, {Peng}, {Barth}  \& {Im}}{{Kim}
  et~al.}{2008}]{kim08}
{Kim} M.,  {Ho} L.~C.,  {Peng} C.~Y.,  {Barth} A.~J.,   {Im} M.,  2008, \mn@doi
  [\apjs] {10.1086/591796}, \href
  {http://adsabs.harvard.edu/abs/2008ApJS..179..283K} {179, 283}

\bibitem[\protect\citeauthoryear{{King}}{{King}}{2003}]{king03}
{King} A.,  2003, \mn@doi [\apjl] {10.1086/379143}, \href
  {http://adsabs.harvard.edu/abs/2003ApJ...596L..27K} {596, L27}

\bibitem[\protect\citeauthoryear{{Knapen}, {Shlosman}  \& {Peletier}}{{Knapen}
  et~al.}{2000}]{knapen00}
{Knapen} J.~H.,  {Shlosman} I.,   {Peletier} R.~F.,  2000, \mn@doi [\apj]
  {10.1086/308266}, \href {http://adsabs.harvard.edu/abs/2000ApJ...529...93K}
  {529, 93}

\bibitem[\protect\citeauthoryear{{Kocevski} et~al.,}{{Kocevski}
  et~al.}{2012}]{kocevski12}
{Kocevski} D.~D.,  et~al., 2012, \mn@doi [\apj] {10.1088/0004-637X/744/2/148},
  \href {http://adsabs.harvard.edu/abs/2012ApJ...744..148K} {744, 148}

\bibitem[\protect\citeauthoryear{{Kormendy} \& {Bender}}{{Kormendy} \&
  {Bender}}{2011}]{kormendy11b}
{Kormendy} J.,  {Bender} R.,  2011, \mn@doi [\nat] {10.1038/nature09695}, \href
  {http://adsabs.harvard.edu/abs/2011Natur.469..377K} {469, 377}

\bibitem[\protect\citeauthoryear{{Kormendy} \& {Gebhardt}}{{Kormendy} \&
  {Gebhardt}}{2001}]{kormendy01}
{Kormendy} J.,  {Gebhardt} K.,  2001, in AIP Conf. Proc. 586: 20th Texas
  Symposium on relativistic astrophysics. pp 363--+

\bibitem[\protect\citeauthoryear{{Kormendy} \& {Ho}}{{Kormendy} \&
  {Ho}}{2013}]{kormendy13}
{Kormendy} J.,  {Ho} L.~C.,  2013, \mn@doi [\araa]
  {10.1146/annurev-astro-082708-101811}, \href
  {http://adsabs.harvard.edu/abs/2013ARA%26A..51..511K} {51, 511}

\bibitem[\protect\citeauthoryear{{Kormendy} \& {Kennicutt}}{{Kormendy} \&
  {Kennicutt}}{2004}]{kormendy04}
{Kormendy} J.,  {Kennicutt} Jr. R.~C.,  2004, \mn@doi [\araa]
  {10.1146/annurev.astro.42.053102.134024}, \href
  {http://adsabs.harvard.edu/abs/2004ARA%26A..42..603K} {42, 603}

\bibitem[\protect\citeauthoryear{{Kormendy}, {Bender}  \& {Cornell}}{{Kormendy}
  et~al.}{2011}]{kormendy11a}
{Kormendy} J.,  {Bender} R.,   {Cornell} M.~E.,  2011, \mn@doi [\nat]
  {10.1038/nature09694}, \href
  {http://adsabs.harvard.edu/abs/2011Natur.469..374K} {469, 374}

\bibitem[\protect\citeauthoryear{{Koss}, {Mushotzky}, {Veilleux}, {Winter},
  {Baumgartner}, {Tueller}, {Gehrels}  \& {Valencic}}{{Koss}
  et~al.}{2011}]{koss11}
{Koss} M.,  {Mushotzky} R.,  {Veilleux} S.,  {Winter} L.~M.,  {Baumgartner} W.,
   {Tueller} J.,  {Gehrels} N.,   {Valencic} L.,  2011, \mn@doi [\apj]
  {10.1088/0004-637X/739/2/57}, \href
  {http://adsabs.harvard.edu/abs/2011ApJ...739...57K} {739, 57}

\bibitem[\protect\citeauthoryear{{Lin}, {Wang}, {Hsieh}, {Taam}, {Yang}  \&
  {Yen}}{{Lin} et~al.}{2013}]{lin13}
{Lin} L.-H.,  {Wang} H.-H.,  {Hsieh} P.-Y.,  {Taam} R.~E.,  {Yang} C.-C.,
  {Yen} D.~C.~C.,  2013, \mn@doi [\apj] {10.1088/0004-637X/771/1/8}, \href
  {http://adsabs.harvard.edu/abs/2013ApJ...771....8L} {771, 8}

\bibitem[\protect\citeauthoryear{{Lintott} et~al.,}{{Lintott}
  et~al.}{2008}]{lintott08}
{Lintott} C.~J.,  et~al., 2008, \mn@doi [\mnras]
  {10.1111/j.1365-2966.2008.13689.x}, \href
  {http://adsabs.harvard.edu/abs/2008MNRAS.389.1179L} {389, 1179}

\bibitem[\protect\citeauthoryear{{Lintott} et~al.,}{{Lintott}
  et~al.}{2011}]{lintott11}
{Lintott} C.,  et~al., 2011, \mn@doi [\mnras]
  {10.1111/j.1365-2966.2010.17432.x}, \href
  {http://adsabs.harvard.edu/abs/2011MNRAS.410..166L} {410, 166}

\bibitem[\protect\citeauthoryear{{Maciejewski}, {Teuben}, {Sparke}  \&
  {Stone}}{{Maciejewski} et~al.}{2002}]{maciejewski02a}
{Maciejewski} W.,  {Teuben} P.~J.,  {Sparke} L.~S.,   {Stone} J.~M.,  2002,
  \mn@doi [\mnras] {10.1046/j.1365-8711.2002.04957.x}, \href
  {http://adsabs.harvard.edu/abs/2002MNRAS.329..502M} {329, 502}

\bibitem[\protect\citeauthoryear{{Magorrian} et~al.,}{{Magorrian}
  et~al.}{1998}]{magorrian98}
{Magorrian} J.,  et~al., 1998, \mn@doi [\aj] {10.1086/300353}, \href
  {http://adsabs.harvard.edu/cgi-bin/nph-bib_query?bibcode=1998AJ....115.2285M&db_key=AST}
  {115, 2285}

\bibitem[\protect\citeauthoryear{{Marconi} \& {Hunt}}{{Marconi} \&
  {Hunt}}{2003}]{marconi03}
{Marconi} A.,  {Hunt} L.~K.,  2003, \mn@doi [\apjl] {10.1086/375804}, \href
  {http://adsabs.harvard.edu/abs/2003ApJ...589L..21M} {589, L21}

\bibitem[\protect\citeauthoryear{{Marleau}, {Clancy}  \& {Bianconi}}{{Marleau}
  et~al.}{2013}]{marleau13}
{Marleau} F.~R.,  {Clancy} D.,   {Bianconi} M.,  2013, \mn@doi [\mnras]
  {10.1093/mnras/stt1503}, \href
  {http://adsabs.harvard.edu/abs/2013MNRAS.435.3085M} {435, 3085}

\bibitem[\protect\citeauthoryear{{Martig}, {Bournaud}, {Croton}, {Dekel}  \&
  {Teyssier}}{{Martig} et~al.}{2012}]{martig12}
{Martig} M.,  {Bournaud} F.,  {Croton} D.~J.,  {Dekel} A.,   {Teyssier} R.,
  2012, \mn@doi [\apj] {10.1088/0004-637X/756/1/26}, \href
  {http://adsabs.harvard.edu/abs/2012ApJ...756...26M} {756, 26}

\bibitem[\protect\citeauthoryear{{Mathur}, {Fields}, {Peterson}  \&
  {Grupe}}{{Mathur} et~al.}{2012}]{mathur12}
{Mathur} S.,  {Fields} D.,  {Peterson} B.~M.,   {Grupe} D.,  2012, \mn@doi
  [\apj] {10.1088/0004-637X/754/2/146}, \href
  {http://adsabs.harvard.edu/abs/2012ApJ...754..146M} {754, 146}

\bibitem[\protect\citeauthoryear{{McConnell} \& {Ma}}{{McConnell} \&
  {Ma}}{2013}]{mcconnell13}
{McConnell} N.~J.,  {Ma} C.-P.,  2013, \mn@doi [\apj]
  {10.1088/0004-637X/764/2/184}, \href
  {http://adsabs.harvard.edu/abs/2013ApJ...764..184M} {764, 184}

\bibitem[\protect\citeauthoryear{{McLure} \& {Dunlop}}{{McLure} \&
  {Dunlop}}{2001}]{mclure01}
{McLure} R.~J.,  {Dunlop} J.~S.,  2001, \mn@doi [\mnras]
  {10.1046/j.1365-8711.2001.04709.x}, \href
  {http://adsabs.harvard.edu/abs/2001MNRAS.327..199M} {327, 199}

\bibitem[\protect\citeauthoryear{{McLure}, {Kukula}, {Dunlop}, {Baum}, {O'Dea}
  \& {Hughes}}{{McLure} et~al.}{1999}]{mclure99}
{McLure} R.~J.,  {Kukula} M.~J.,  {Dunlop} J.~S.,  {Baum} S.~A.,  {O'Dea}
  C.~P.,   {Hughes} D.~H.,  1999, \mnras, \href
  {http://adsabs.harvard.edu/cgi-bin/nph-bib_query?bibcode=1999MNRAS.308..377M&db_key=AST}
  {308, 377}

\bibitem[\protect\citeauthoryear{{Mechtley} et~al.,}{{Mechtley}
  et~al.}{2016}]{mechtley16}
{Mechtley} M.,  et~al., 2016, \mn@doi [\apj] {10.3847/0004-637X/830/2/156},
  \href {http://adsabs.harvard.edu/abs/2016ApJ...830..156M} {830, 156}

\bibitem[\protect\citeauthoryear{{Oh}, {Oh}  \& {Yi}}{{Oh} et~al.}{2012}]{oh12}
{Oh} S.,  {Oh} K.,   {Yi} S.~K.,  2012, \mn@doi [\apjs]
  {10.1088/0067-0049/198/1/4}, \href
  {http://adsabs.harvard.edu/abs/2012ApJS..198....4O} {198, 4}

\bibitem[\protect\citeauthoryear{{Peng}}{{Peng}}{2007}]{peng07}
{Peng} C.~Y.,  2007, \mn@doi [\apj] {10.1086/522774}, \href
  {http://adsabs.harvard.edu/abs/2007ApJ...671.1098P} {671, 1098}

\bibitem[\protect\citeauthoryear{{Peng}, {Ho}, {Impey}  \& {Rix}}{{Peng}
  et~al.}{2002}]{peng02}
{Peng} C.~Y.,  {Ho} L.~C.,  {Impey} C.~D.,   {Rix} H.-W.,  2002, \mn@doi [\aj]
  {10.1086/340952}, \href
  {http://adsabs.harvard.edu/cgi-bin/nph-bib_query?bibcode=2002AJ....124..266P&db_key=AST}
  {124, 266}

\bibitem[\protect\citeauthoryear{{Peng}, {Ho}, {Impey}  \& {Rix}}{{Peng}
  et~al.}{2010}]{peng10}
{Peng} C.~Y.,  {Ho} L.~C.,  {Impey} C.~D.,   {Rix} H.-W.,  2010, \mn@doi [\aj]
  {10.1088/0004-6256/139/6/2097}, \href
  {http://adsabs.harvard.edu/abs/2010AJ....139.2097P} {139, 2097}

\bibitem[\protect\citeauthoryear{{Pierce} et~al.,}{{Pierce}
  et~al.}{2007}]{pierce07}
{Pierce} C.~M.,  et~al., 2007, \mn@doi [\apjl] {10.1086/517922}, \href
  {http://adsabs.harvard.edu/abs/2007ApJ...660L..19P} {660, L19}

\bibitem[\protect\citeauthoryear{{Regan} \& {Teuben}}{{Regan} \&
  {Teuben}}{2004}]{regan04}
{Regan} M.~W.,  {Teuben} P.~J.,  2004, \mn@doi [\apj] {10.1086/380116}, \href
  {http://adsabs.harvard.edu/abs/2004ApJ...600..595R} {600, 595}

\bibitem[\protect\citeauthoryear{{Richards} et~al.,}{{Richards}
  et~al.}{2002}]{richards02}
{Richards} G.~T.,  et~al., 2002, \mn@doi [\aj] {10.1086/340187}, \href
  {http://adsabs.harvard.edu/abs/2002AJ....123.2945R} {123, 2945}

\bibitem[\protect\citeauthoryear{{Richards} et~al.,}{{Richards}
  et~al.}{2006}]{richards06}
{Richards} G.~T.,  et~al., 2006, \mn@doi [\apjs] {10.1086/506525}, \href
  {http://adsabs.harvard.edu/abs/2006ApJS..166..470R} {166, 470}

\bibitem[\protect\citeauthoryear{{S{\' a}nchez} et~al.,}{{S{\' a}nchez}
  et~al.}{2004}]{sanchez04}
{S{\' a}nchez} S.~F.,  et~al., 2004, \mn@doi [\apj] {10.1086/423234}, \href
  {http://adsabs.harvard.edu/cgi-bin/nph-bib_query?bibcode=2004ApJ...614..586S&db_key=AST}
  {614, 586}

\bibitem[\protect\citeauthoryear{{Sakamoto}}{{Sakamoto}}{1996}]{sakamoto96}
{Sakamoto} K.,  1996, \mn@doi [\apj] {10.1086/177961}, \href
  {http://adsabs.harvard.edu/abs/1996ApJ...471..173S} {471, 173}

\bibitem[\protect\citeauthoryear{{Sanders}, {Soifer}, {Elias}, {Madore},
  {Matthews}, {Neugebauer}  \& {Scoville}}{{Sanders} et~al.}{1988}]{sanders88}
{Sanders} D.~B.,  {Soifer} B.~T.,  {Elias} J.~H.,  {Madore} B.~F.,  {Matthews}
  K.,  {Neugebauer} G.,   {Scoville} N.~Z.,  1988, \mn@doi [\apj]
  {10.1086/165983}, \href {http://adsabs.harvard.edu/abs/1988ApJ...325...74S}
  {325, 74}

\bibitem[\protect\citeauthoryear{{Sarzi} et~al.,}{{Sarzi}
  et~al.}{2006}]{sarzi06}
{Sarzi} M.,  et~al., 2006, \mn@doi [\mnras] {10.1111/j.1365-2966.2005.09839.x},
  \href {http://adsabs.harvard.edu/abs/2006MNRAS.366.1151S} {366, 1151}

\bibitem[\protect\citeauthoryear{{Satyapal}, {B{\"o}ker}, {Mcalpine},
  {Gliozzi}, {Abel}  \& {Heckman}}{{Satyapal} et~al.}{2009}]{satyapal09}
{Satyapal} S.,  {B{\"o}ker} T.,  {Mcalpine} W.,  {Gliozzi} M.,  {Abel} N.~P.,
  {Heckman} T.,  2009, \mn@doi [\apj] {10.1088/0004-637X/704/1/439}, \href
  {http://adsabs.harvard.edu/abs/2009ApJ...704..439S} {704, 439}

\bibitem[\protect\citeauthoryear{{Satyapal}, {Secrest}, {McAlpine}, {Ellison},
  {Fischer}  \& {Rosenberg}}{{Satyapal} et~al.}{2014}]{satyapal14}
{Satyapal} S.,  {Secrest} N.~J.,  {McAlpine} W.,  {Ellison} S.~L.,  {Fischer}
  J.,   {Rosenberg} J.~L.,  2014, \mn@doi [\apj] {10.1088/0004-637X/784/2/113},
  \href {http://adsabs.harvard.edu/abs/2014ApJ...784..113S} {784, 113}

\bibitem[\protect\citeauthoryear{{Schawinski}, {Treister}, {Urry}, {Cardamone},
  {Simmons}  \& {Yi}}{{Schawinski} et~al.}{2011a}]{schawinski11a}
{Schawinski} K.,  {Treister} E.,  {Urry} C.~M.,  {Cardamone} C.~N.,  {Simmons}
  B.,   {Yi} S.~K.,  2011a, \mn@doi [\apjl] {10.1088/2041-8205/727/2/L31},
  \href {http://adsabs.harvard.edu/abs/2011ApJ...727L..31S} {727, L31+}

\bibitem[\protect\citeauthoryear{{Schawinski}, {Urry}, {Treister}, {Simmons},
  {Natarajan}  \& {Glikman}}{{Schawinski} et~al.}{2011b}]{schawinski11b}
{Schawinski} K.,  {Urry} M.,  {Treister} E.,  {Simmons} B.,  {Natarajan} P.,
  {Glikman} E.,  2011b, \mn@doi [\apjl] {10.1088/2041-8205/743/2/L37}, \href
  {http://adsabs.harvard.edu/abs/2011ApJ...743L..37S} {743, L37}

\bibitem[\protect\citeauthoryear{{Schawinski}, {Simmons}, {Urry}, {Treister}
  \& {Glikman}}{{Schawinski} et~al.}{2012}]{schawinski12}
{Schawinski} K.,  {Simmons} B.~D.,  {Urry} C.~M.,  {Treister} E.,   {Glikman}
  E.,  2012, \mn@doi [\mnras] {10.1111/j.1745-3933.2012.01302.x}, \href
  {http://adsabs.harvard.edu/abs/2012MNRAS.425L..61S} {425, L61}

\bibitem[\protect\citeauthoryear{{Schawinski}, {Koss}, {Berney}  \&
  {Sartori}}{{Schawinski} et~al.}{2015}]{schawinski15}
{Schawinski} K.,  {Koss} M.,  {Berney} S.,   {Sartori} L.~F.,  2015, \mn@doi
  [\mnras] {10.1093/mnras/stv1136}, \href
  {http://adsabs.harvard.edu/abs/2015MNRAS.451.2517S} {451, 2517}

\bibitem[\protect\citeauthoryear{{Schlegel}, {Finkbeiner}  \&
  {Davis}}{{Schlegel} et~al.}{1998}]{schlegel98}
{Schlegel} D.~J.,  {Finkbeiner} D.~P.,   {Davis} M.,  1998, \mn@doi [\apj]
  {10.1086/305772}, \href {http://adsabs.harvard.edu/abs/1998ApJ...500..525S}
  {500, 525}

\bibitem[\protect\citeauthoryear{{Secrest}, {Satyapal}, {Gliozzi}, {Cheung},
  {Seth}  \& {B{\"o}ker}}{{Secrest} et~al.}{2012}]{secrest12}
{Secrest} N.~J.,  {Satyapal} S.,  {Gliozzi} M.,  {Cheung} C.~C.,  {Seth} A.~C.,
    {B{\"o}ker} T.,  2012, \mn@doi [\apj] {10.1088/0004-637X/753/1/38}, \href
  {http://adsabs.harvard.edu/abs/2012ApJ...753...38S} {753, 38}

\bibitem[\protect\citeauthoryear{{Secrest}, {Satyapal}, {Moran}, {Cheung},
  {Giroletti}, {Gliozzi}, {Bergmann}  \& {Seth}}{{Secrest}
  et~al.}{2013}]{secrest13}
{Secrest} N.~J.,  {Satyapal} S.,  {Moran} S.~M.,  {Cheung} C.~C.,  {Giroletti}
  M.,  {Gliozzi} M.,  {Bergmann} M.~P.,   {Seth} A.~C.,  2013, \mn@doi [\apj]
  {10.1088/0004-637X/777/2/139}, \href
  {http://adsabs.harvard.edu/abs/2013ApJ...777..139S} {777, 139}

\bibitem[\protect\citeauthoryear{{Shen} et~al.,}{{Shen} et~al.}{2011}]{shen11}
{Shen} Y.,  et~al., 2011, \mn@doi [\apjs] {10.1088/0067-0049/194/2/45}, \href
  {http://adsabs.harvard.edu/abs/2011ApJS..194...45S} {194, 45}

\bibitem[\protect\citeauthoryear{{Silk} \& {Rees}}{{Silk} \&
  {Rees}}{1998}]{silk98}
{Silk} J.,  {Rees} M.~J.,  1998, \aap, \href
  {http://adsabs.harvard.edu/abs/1998A%26A...331L...1S} {331, L1}

\bibitem[\protect\citeauthoryear{{Simard}}{{Simard}}{1998}]{simard98}
{Simard} L.,  1998, in {Albrecht} R.,  {Hook} R.~N.,   {Bushouse} H.~A.,  eds,
  Astronomical Society of the Pacific Conference Series Vol. 145, Astronomical
  Data Analysis Software and Systems VII. p.~108

\bibitem[\protect\citeauthoryear{{Simard}, {Mendel}, {Patton}, {Ellison}  \&
  {McConnachie}}{{Simard} et~al.}{2011}]{simard11}
{Simard} L.,  {Mendel} J.~T.,  {Patton} D.~R.,  {Ellison} S.~L.,
  {McConnachie} A.~W.,  2011, \mn@doi [\apjs] {10.1088/0067-0049/196/1/11},
  \href {http://adsabs.harvard.edu/abs/2011ApJS..196...11S} {196, 11}

\bibitem[\protect\citeauthoryear{{Simmons} \& {Urry}}{{Simmons} \&
  {Urry}}{2008}]{simmons08}
{Simmons} B.~D.,  {Urry} C.~M.,  2008, \mn@doi [\apj] {10.1086/589827}, \href
  {http://adsabs.harvard.edu/abs/2008ApJ...683..644S} {683, 644}

\bibitem[\protect\citeauthoryear{{Simmons}, {Van Duyne}, {Urry}, {Treister},
  {Koekemoer}, {Grogin}  \& {The GOODS Team}}{{Simmons}
  et~al.}{2011}]{simmons11}
{Simmons} B.~D.,  {Van Duyne} J.,  {Urry} C.~M.,  {Treister} E.,  {Koekemoer}
  A.~M.,  {Grogin} N.~A.,   {The GOODS Team} 2011, \mn@doi [\apj]
  {10.1088/0004-637X/734/2/121}, \href
  {http://adsabs.harvard.edu/abs/2011ApJ...734..121S} {734, 121}

\bibitem[\protect\citeauthoryear{{Simmons}, {Urry}, {Schawinski}, {Cardamone}
  \& {Glikman}}{{Simmons} et~al.}{2012}]{simmons12b}
{Simmons} B.~D.,  {Urry} C.~M.,  {Schawinski} K.,  {Cardamone} C.,   {Glikman}
  E.,  2012, \mn@doi [\apj] {10.1088/0004-637X/761/1/75}, \href
  {http://adsabs.harvard.edu/abs/2012ApJ...761...75S} {761, 75}

\bibitem[\protect\citeauthoryear{{Simmons} et~al.,}{{Simmons}
  et~al.}{2013}]{simmons13}
{Simmons} B.~D.,  et~al., 2013, \mn@doi [\mnras] {10.1093/mnras/sts491}, \href
  {http://adsabs.harvard.edu/abs/2013MNRAS.429.2199S} {429, 2199}

\bibitem[\protect\citeauthoryear{{Simpson} et~al.,}{{Simpson}
  et~al.}{2013}]{rsimpson13}
{Simpson} R.~J.,  et~al., 2013, preprint, \href
  {http://adsabs.harvard.edu/abs/2013arXiv1301.5193S} {} (\mn@eprint {arXiv}
  {1301.5193})

\bibitem[\protect\citeauthoryear{{Skrutskie} et~al.,}{{Skrutskie}
  et~al.}{2006}]{skrutskie06}
{Skrutskie} M.~F.,  et~al., 2006, \mn@doi [\aj] {10.1086/498708}, \href
  {http://adsabs.harvard.edu/abs/2006AJ....131.1163S} {131, 1163}

\bibitem[\protect\citeauthoryear{{Stern} et~al.,}{{Stern}
  et~al.}{2005}]{dstern05}
{Stern} D.,  et~al., 2005, \mn@doi [\apj] {10.1086/432523}, \href
  {http://adsabs.harvard.edu/abs/2005ApJ...631..163S} {631, 163}

\bibitem[\protect\citeauthoryear{{Taylor}}{{Taylor}}{2005}]{Taylor05}
{Taylor} M.~B.,  2005, in {Shopbell} P.,  {Britton} M.,   {Ebert} R.,  eds,
  Astronomical Society of the Pacific Conference Series Vol. 347, Astronomical
  Data Analysis Software and Systems XIV. p.~29

\bibitem[\protect\citeauthoryear{{Toomre}}{{Toomre}}{1977}]{toomre77}
{Toomre} A.,  1977, in {B.~M.~Tinsley \& R.~B.~G.~Larson D.~Campbell} ed.,
  Evolution of Galaxies and Stellar Populations. p.~401

\bibitem[\protect\citeauthoryear{{Tortora}, {Antonuccio-Delogu}, {Kaviraj},
  {Silk}, {Romeo}  \& {Becciani}}{{Tortora} et~al.}{2009}]{tortora09}
{Tortora} C.,  {Antonuccio-Delogu} V.,  {Kaviraj} S.,  {Silk} J.,  {Romeo}
  A.~D.,   {Becciani} U.,  2009, \mn@doi [\mnras]
  {10.1111/j.1365-2966.2009.14718.x}, \href
  {http://adsabs.harvard.edu/abs/2009MNRAS.396...61T} {396, 61}

\bibitem[\protect\citeauthoryear{{Trakhtenbrot}, {Lira}, {Netzer}, {Cicone},
  {Maiolino}  \& {Shemmer}}{{Trakhtenbrot} et~al.}{2016}]{trakhtenbrot16b}
{Trakhtenbrot} B.,  {Lira} P.,  {Netzer} H.,  {Cicone} C.,  {Maiolino} R.,
  {Shemmer} O.,  2016, preprint, \href
  {http://adsabs.harvard.edu/abs/2016arXiv161200010T} {} (\mn@eprint {arXiv}
  {1612.00010})

\bibitem[\protect\citeauthoryear{{Urrutia}, {Lacy}  \& {Becker}}{{Urrutia}
  et~al.}{2008}]{urrutia08}
{Urrutia} T.,  {Lacy} M.,   {Becker} R.~H.,  2008, \mn@doi [\apj]
  {10.1086/523959}, \href {http://adsabs.harvard.edu/abs/2008ApJ...674...80U}
  {674, 80}

\bibitem[\protect\citeauthoryear{{Urry}, {Scarpa}, {O'Dowd}, {Falomo}, {Pesce}
  \& {Treves}}{{Urry} et~al.}{2000}]{urry00}
{Urry} C.~M.,  {Scarpa} R.,  {O'Dowd} M.,  {Falomo} R.,  {Pesce} J.~E.,
  {Treves} A.,  2000, \mn@doi [\apj] {10.1086/308616}, \href
  {http://adsabs.harvard.edu/cgi-bin/nph-bib_query?bibcode=2000ApJ...532..816U&db_key=AST}
  {532, 816}

\bibitem[\protect\citeauthoryear{{Voges} et~al.,}{{Voges}
  et~al.}{1999}]{voges99}
{Voges} W.,  et~al., 1999, \aap, \href
  {http://adsabs.harvard.edu/abs/1999A%26A...349..389V} {349, 389}

\bibitem[\protect\citeauthoryear{{Volonteri}}{{Volonteri}}{2010}]{volonteri10}
{Volonteri} M.,  2010, \mn@doi [\aapr] {10.1007/s00159-010-0029-x}, \href
  {http://adsabs.harvard.edu/abs/2010A%26ARv..18..279V} {18, 279}

\bibitem[\protect\citeauthoryear{{Volonteri}, {Lodato}  \&
  {Natarajan}}{{Volonteri} et~al.}{2008}]{volonteri08}
{Volonteri} M.,  {Lodato} G.,   {Natarajan} P.,  2008, \mn@doi [\mnras]
  {10.1111/j.1365-2966.2007.12589.x}, \href
  {http://adsabs.harvard.edu/abs/2008MNRAS.383.1079V} {383, 1079}

\bibitem[\protect\citeauthoryear{{Volonteri}, {Natarajan}  \&
  {G{\"u}ltekin}}{{Volonteri} et~al.}{2011}]{volonteri11}
{Volonteri} M.,  {Natarajan} P.,   {G{\"u}ltekin} K.,  2011, \mn@doi [\apj]
  {10.1088/0004-637X/737/2/50}, \href
  {http://adsabs.harvard.edu/abs/2011ApJ...737...50V} {737, 50}

\bibitem[\protect\citeauthoryear{{Walker}, {Mihos}  \& {Hernquist}}{{Walker}
  et~al.}{1996}]{walker96}
{Walker} I.~R.,  {Mihos} J.~C.,   {Hernquist} L.,  1996, \mn@doi [\apj]
  {10.1086/176956}, \href {http://adsabs.harvard.edu/abs/1996ApJ...460..121W}
  {460, 121}

\bibitem[\protect\citeauthoryear{{Welker}, {Dubois}, {Devriendt}, {Pichon},
  {Kaviraj}  \& {Peirani}}{{Welker} et~al.}{2015}]{welker15}
{Welker} C.,  {Dubois} Y.,  {Devriendt} J.,  {Pichon} C.,  {Kaviraj} S.,
  {Peirani} S.,  2015, preprint, \href
  {http://adsabs.harvard.edu/abs/2015arXiv150205053W} {} (\mn@eprint {arXiv}
  {1502.05053})

\bibitem[\protect\citeauthoryear{{Wright}}{{Wright}}{2006}]{wright06}
{Wright} E.~L.,  2006, \mn@doi [\pasp] {10.1086/510102}, \href
  {http://adsabs.harvard.edu/abs/2006PASP..118.1711W} {118, 1711}

\bibitem[\protect\citeauthoryear{{Wright} et~al.,}{{Wright}
  et~al.}{2010}]{wright10}
{Wright} E.~L.,  et~al., 2010, \mn@doi [\aj] {10.1088/0004-6256/140/6/1868},
  \href {http://adsabs.harvard.edu/abs/2010AJ....140.1868W} {140, 1868}

\bibitem[\protect\citeauthoryear{{Xiao}, {Barth}, {Greene}, {Ho}, {Bentz},
  {Ludwig}  \& {Jiang}}{{Xiao} et~al.}{2011}]{xiao11}
{Xiao} T.,  {Barth} A.~J.,  {Greene} J.~E.,  {Ho} L.~C.,  {Bentz} M.~C.,
  {Ludwig} R.~R.,   {Jiang} Y.,  2011, \mn@doi [\apj]
  {10.1088/0004-637X/739/1/28}, \href
  {http://adsabs.harvard.edu/abs/2011ApJ...739...28X} {739, 28}

\bibitem[\protect\citeauthoryear{{York} et~al.,}{{York} et~al.}{2000}]{york00}
{York} D.~G.,  et~al., 2000, \mn@doi [\aj] {10.1086/301513}, \href
  {http://adsabs.harvard.edu/abs/2000AJ....120.1579Y} {120, 1579}

\bibitem[\protect\citeauthoryear{{de Jong} \& {Bell}}{{de Jong} \&
  {Bell}}{2007}]{dejong07}
{de Jong} R.~S.,  {Bell} E.~F.,  2007, \mn@doi [Astrophysics and Space Science
  Proceedings] {10.1007/978-1-4020-5573-7_16}, \href
  {http://adsabs.harvard.edu/abs/2007ASSP....3..107D} {3, 107}

\bibitem[\protect\citeauthoryear{{de Vaucouleurs}}{{de
  Vaucouleurs}}{1953}]{devaucouleurs}
{de Vaucouleurs} G.,  1953, \mnras, \href
  {http://adsabs.harvard.edu/abs/1953MNRAS.113..134D} {113, 134}

\bibitem[\protect\citeauthoryear{{van den Bosch}}{{van den
  Bosch}}{2016}]{vandenbosch16}
{van den Bosch} R.~C.~E.,  2016, \mn@doi [\apj] {10.3847/0004-637X/831/2/134},
  \href {http://adsabs.harvard.edu/abs/2016ApJ...831..134V} {831, 134}

\makeatother
\end{thebibliography}

\bsp	
\label{lastpage}
\end{document}